\documentclass[graybox]{svmult}

\usepackage{mathptmx}       % selects Times Roman as basic font
\usepackage{helvet}         % selects Helvetica as sans-serif font
\usepackage{courier}        % selects Courier as typewriter font
\usepackage{type1cm}        % activate if the above 3 fonts are
\usepackage{makeidx}         % allows index generation
\usepackage{graphicx}        % standard LaTeX graphics tool
\usepackage{multicol}        % used for the two-column index
\usepackage[bottom]{footmisc}% places footnotes at page bottom
\usepackage{cite}
\usepackage{wrapfig}
\usepackage{sidecap}
\usepackage{amssymb,amsmath}

\usepackage{hyperref}
\usepackage{epstopdf}

\makeindex             % used for the subject index
\newcommand{\re}[1]{\mathrm{Re}\,#1}
\newcommand{\im}[1]{\mathrm{Im}\,#1}
\renewcommand{\imath}[0]{\mathrm{i}}
\newcommand{\msum}{\sideset{}{'}\sum_{n=0}^{\infty}}

\usepackage{color}

\begin{document}

\title*{Fluctuation-induced forces between atoms and surfaces: the Casimir-Polder 
interaction}
\author{Francesco Intravaia, Carsten Henkel, and Mauro Antezza}
\institute{F. Intravaia \at Theoretical Division, MS B213, Los Alamos National 
Laboratory, Los Alamos, NM87545, USA \email{francesco\_intravaia@lanl.gov}
\and C. Henkel \at Institut fuer Physik und Astronomie, Universitaet Potsdam, 
14476 Potsdam, Germany, \email{carsten.henkel@physik.uni-potsdam.de}
\and M.Antezza \at Laboratoire Kastler Brossel, ƒcole Normale SupŽrieure, CNRS 
and UPMC, 24 rue Lhomond, 75231 Paris, France \email{mauro.antezza@lkb.ens.fr}
}
\maketitle

\abstract{
%{}[compiled \today]
Electromagnetic fluctuation-induced forces between atoms and surfaces are generally known as  Casimir-Polder interactions. The exact knowledge of these forces is rapidly becoming important in modern experimental set-ups and for technological applications. Recent theoretical and experimental investigations have 
shown that such an interaction is tunable in strength and sign, opening new 
perspectives to investigate aspects of quantum field theory and 
condensed-matter physics. In this Chapter we review the theory of fluctuation-%
induced interactions between atoms and a surface, paying particular attention to 
the physical characterization of the system. We also survey some recent 
developments concerning the role of temperature, situations out of thermal
equilibrium, and measurements involving ultra-cold atoms.
}

\section{Introduction}

In the last decade remarkable progress in trapping and manipulating atoms has 
opened a wide horizon to new and challenging experimental set-ups. 
Precision tests of both quantum mechanics and quantum electrodynamics
have become possible through the capacity of addressing single trapped
particles \cite{Diedrich89,Meekhof96} and of cooling ultracold gases down 
to Bose-Einstein condensation \cite{PitaevskiiStringari,Antezza04,Obrecht07}.
This stunning progress is also very profitable to other fundamental areas of physics and to technology.
For example, ultracold gases have been suggested as probes in interesting experimental proposals aiming at very accurate tests of the gravity law \cite{Carusotto05,Wolf07,Sorrentino09}, looking for extra forces predicted by 
different grand-unified theories \cite{Onofrio06}
(see also the chapter by Milton in this volume for detailed discussions on the interplay between Casimir energy and Gravity).
Technologically speaking, one paradigmatic example of 
this new frontier is provided by atom chips \cite{Folman02,Fortagh07}. 
In these tiny devices, a cloud of atoms 
(typically alkalis like Sodium, Rubidium or Cesium) is magnetically or optically 
trapped above a patterned surface, reaching relatively short 
distances between a few microns to hundreds of microns \cite{Reichel99,Folman00,Zimmermann01a}. The micro-machined surface
patterns form a system of conducting wires, which are used to control the 
atomic cloud by tuning an external induced current (also superconducting 
wires have 
been demonstrated \cite{Roux08, Cano08b, Muller09a}).

\begin{figure}[htb]
\begin{center}
\includegraphics*[width=8cm]{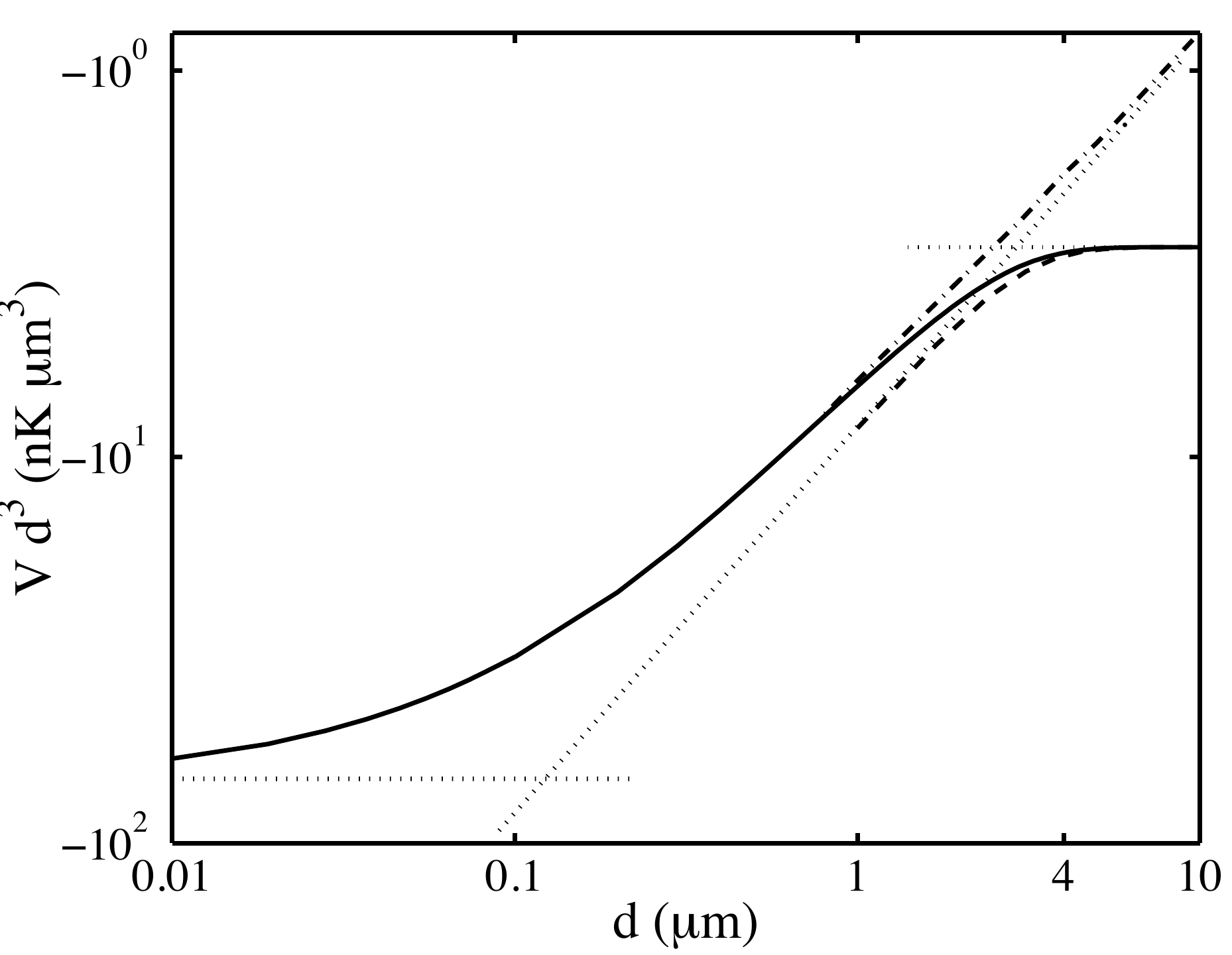}
\end{center}
\caption{Atom-surface potential 
(free energy of interaction $V(d)$) vs.\ distance $d$
between a $^{87}$Rb atom and a SiO$_2$ 
(sapphire) substrate, multiplied by $d^3$.  The potential is 
calculated using the theory of Dzyaloshinskii, Lifshitz,
and Pitaevskii \cite[Ch.VIII]{Landau9}. Note the logarithmic scale and
the sign. The figure, adapted from Fig.3 
of \cite{Antezza04}, shows the potential 
at $T=300\,{\rm K}$ (solid line), at $T=0\,{\rm K}$ 
(dash-dotted line), and the three asymptotic behaviors (dotted lines): van der 
Waals-London~$\propto -1/d^3$, Casimir-Polder ($\propto -1/d^4$), and 
Lifshitz ($\propto -T/d^3$).) 
%The ratio between the short- and large-
%distance asymptotes is determined by the dominant Bohr frequencies 
%$\omega_{\rm Rb}$ in the Rubidium atom, 
%of the order of $k_B T / \hbar \omega_{\rm Rb}$.
}
\label{fig:potential-overview}
\end{figure}

At a fundamental level, all these systems have in common to be strongly 
influenced by all kinds of atom-surface interactions. 
A particular category are  
fluctuation-induced forces, of which the most prominent representative is the van 
der Waals interaction \cite{Parsegian06}. 
These forces usually derive from a potential with a characteristic power-law dependence
\begin{equation}
\mbox{van der Waals limit}: \quad
V = V_{\rm vdW}\propto\frac{1}{L^{n}}
	\label{powerlaw1}~,
\end{equation}
where $L$ is the distance between the objects (two atoms or a surface and a 
atom) and the exponent depends upon physical parameters and geometry of the 
system ($n = 3$ for an atom and a thick plate, see Fig.\ref{fig:potential-overview}).
Historically speaking, the 
existence of this kind of interaction was 
postulated long before it was experimentally possible to address single atoms \cite{Margenau39}. 
The first quantum-mechanical theory was formulated by F. London in the thirties 
using the idea that the quantum mechanical uncertainty of electrons in atoms can
be translated into fluctuating electric dipole moments
\cite{London30}. London found that two atoms attract each other following \eqref
{powerlaw1} with an exponent $n=6$. London's theory was extensively applied 
in studying colloidal suspensions \cite{Verweey48} which provided confirmations of
its validity but also showed its limitations. 

The next step was taken by H.B.G. Casimir and 
his student D. Polder \cite{Casimir48a} who applied the framework of quantum 
electrodynamics, including the concept of vacuum (field) fluctuations. They generalized the London-van der Waals formula by relaxing the electrostatic 
approximation, in other words, including the effect of retardation. 
The main success
of Casimir-Polder theory was to provide an explanation for the change in the 
power law exponent observed in some experiments\cite{Verweey48}. Indeed 
for distances larger than a characteristic length scale $\lambda_{0}$ of the 
system,  the effect of retardation can no longer be neglected, and this leads to
\begin{equation}
\mbox{Casimir-Polder limit $L\gg \lambda_{0}$:} \quad
V = V_{\rm CP} \propto \frac{ \lambda_0 }{ L^{n+1} }
	\label{powerlaw2}~,	
\end{equation}
and therefore to the $L^{-7}$ dependence typical of the Casimir-Polder interaction 
between two atoms. 
The scale $\lambda_{0}$ is in this case the wavelength of the main atomic absorption 
lines, which is in the visible to near infrared for typical alkali atoms, 
a few hundreds of nanometers.

The estimates~(\ref{powerlaw1}, \ref{powerlaw2}) apply at $T = 0$ when only
quantum fluctuations play a role. If the temperature is nonzero, another length
scale comes into play, the thermal or Wien wavelength 
\begin{equation}
	\lambda_{T} = \frac{\hbar\, c}{k_{B}T}~,
\end{equation}
which corresponds to the wave length where the thermal radiation spectrum peaks. 
Calculations of the atom-surface interaction using thermal quantum field theory have 
been pioneered by Dzyaloshinskii, Lifshitz, and Pitaevskii \cite{Landau9,Dzyaloshinskii61}. They were able to recover the van der Waals and the Casimir-Polder potentials as limit behavior of a more general expression and to confirm, quite surprisingly, that at distances $L \ll \lambda_{T}$ the interactions are
typically dominated by quantum fluctuations, the main reason being that their
spectrum is much wider than that of the thermal field (which is constrained by the 
Bose-Einstein distribution \cite{Mandel95}). At distances $L \gg \lambda_T$ they show that the potential shows again a cross-over from the Casimir-Polder to the
Lifshitz asymptote:
\begin{equation}
\mbox{Lifshitz limit $L\gg \lambda_{T}$:} \quad
V = V_{\rm L} \propto \frac{ \lambda_0 }{ \lambda_{T} L^{n} } \sim 
	\frac{ k_B T }{ \hbar \omega_0 } V_{\rm vdW}
	\label{powerlaw3}~,
\end{equation}
where $\omega_0$ is the (angular) frequency corresponding to $\lambda_0$.
This potential that scales with temperature is actually a free energy of interaction
and is also known as the Keesom potential between polar molecules: there, the dipoles 
are rotating freely under the influence of thermal fluctuations \cite{Linder64}.
In this case (Rydberg atoms provide another example), the particle resonances
overlap with the thermal spectrum, and the Casimir-Polder regime is actually 
absent. Eq.(\ref{powerlaw3}) predicts an apparent enhancement, at nonzero
temperature, of the fluctuation-induced interaction. This does not necessarily
happen, however, because
the molecular polarizabilities are also temperature-dependent 
\cite{Linder64,Haakh09b,Ellingsen10a}.

In Fig.\ref{fig:potential-overview} above, we considered the case of an alkali atom
whose peak
absorption wavelength $\lambda_0$ is much shorter than the Wien wavelength.
The Lifshitz tail is then much smaller than the van der Waals potential.
Note that $\lambda_{T}$ is of the order of a few micrometers at room temperature,
comparable to the smallest atom-surface distances achieved so far
in atom chips. The 
crossover between the Casimir--Polder and the Lifshitz regimes can thus
be explored in these set-ups. We discuss corresponding experiments in 
Sec. \ref{sec:Exp}. 

In the following sections, we start with a derivation of the
interaction between an atom and a general electromagnetic
environment 
(Sec.\ref{s:Atom-Surface interaction}). We will refer to it using the term ``atom-surface interaction'' or ``Casimir-Polder interaction''. This latter is also of common use in the literature to stress the fluctuation-induced nature of the interaction, although the term ``Casimir-Polder'' more correctly indicates the potential in the retarded limit (see Eq.\eqref{powerlaw2}).
The result will be valid within a second-order perturbation theory and can 
be easily adapted to specific geometries. We provide some details on
a planar surface (Sec.\ref{s:Green-Tensor-planar-surface}). Situations
out of global thermal equilibrium are discussed in 
Sec.\ref{s:non-equilibrium}, dealing with forces on ultracold atoms in a 
%``hot'' 
%radiation field
general radiation environment (the temperature of the surface and that of the surrounding environment are not necessarily the same), and with radiative friction. The final Sec.\ref{sec:Exp} sketches experiments with atomic beams and ultracold samples.

%==================================================
\section{Understanding atom-surface interactions}
	\label{s:Atom-Surface interaction}
%==================================================

The interaction between atoms and between atoms and surfaces plays a 
fundamental role in many fields of physics, chemistry and technology 
(see also the chapter of DeKieviet {\it et al.} in this volume for detailed discussions on modern experiments on atom-surface Casimir physics).
From a quantum-mechanical point of view, fluctuation-induced forces are not 
surprising and almost a natural consequence of the initial assumptions. Indeed the existence of fluctuations, 
even at zero temperature, is one of the most remarkable predictions of the theory. 
Each observable corresponding to a physically measurable quantity can be zero on 
average but its variance will always be nonzero if the system is not in one of its 
eigenstates. When two quantum systems interact, the dynamics of the fluctuations 
becomes richer: each subsystem experiences, in 
addition to its own fluctuations, an external, fluctuating force.
%Figure
\begin{wrapfigure}{r}{4cm}
 \center
 \includegraphics[width=4cm]{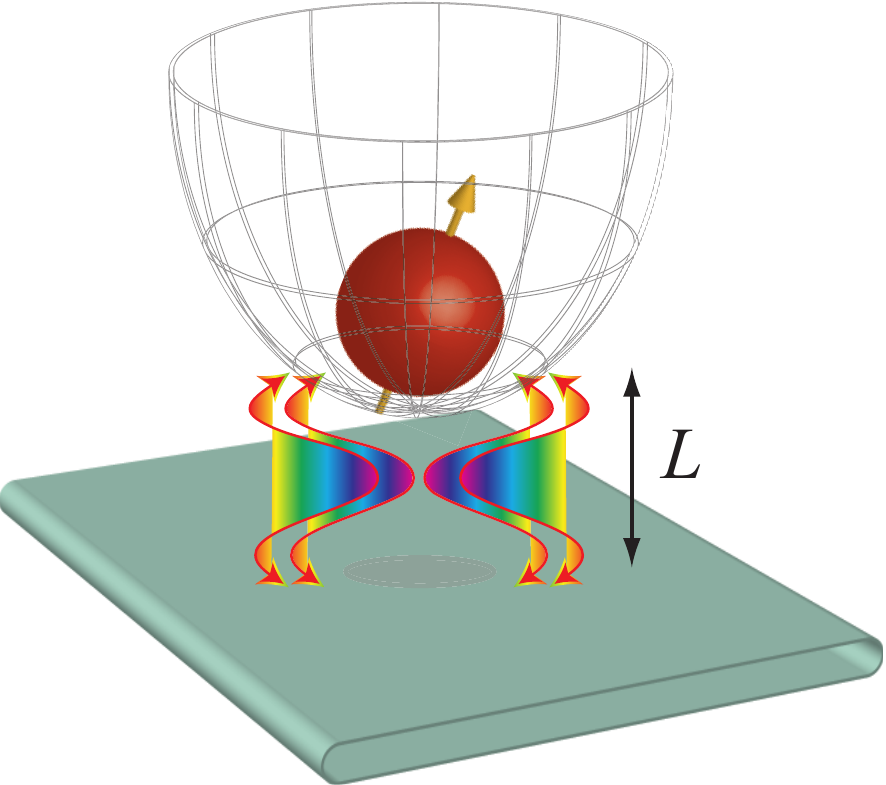}
   \caption{A schematic representation of the atom in the trap near a surface.}
  \label{fig:atomsurface}
\end{wrapfigure} 
%End Figure
This becomes particularly clear in the case of a polarizable 
particle (atom or nanosphere) interacting with the vacuum electromagnetic 
field. In vacuum, the
electromagnetic field fluctuates not only by itself, but also because there are
fluctuating sources for it, like the electric dipole moment of the particle. 
At the same time, 
the particle's dipole is not only fluctuating on its own, but is also responding to
the fluctuations of the electromagnetic field \cite{Dalibard82,Dalibard84}. As a 
result, when two atoms are brought nearby, they interact through their 
fluctuations 
mediated by the electromagnetic field. Similarly when a particle is in proximity of 
a macroscopic object, electric currents fluctuating inside the object and the 
fluctuations 
of the particle lead to a distance-dependent force. This second case is 
complicated by the fact of dealing with a macroscopic object and its 
quantum-mechanical
description. However, if the medium responds linearly to an electromagnetic 
perturbation, the fluctuation-dissipation theorem \cite{Callen51}
% Rytov's theory \cite{Rytov53} 
provides a connection between the field's autocorrelation function and 
its macroscopic response (or Green function).

We will use the previous considerations as a starting point for the derivation of the 
Casimir-Polder interaction between a surface and an atom or also a nanoparticle. 
We will follow Refs\cite{Meschede90,Haakh09b,Henkel02,Novotny08}.  Although 
this is not the unique approach \cite{Casimir48a,Dzyaloshinskii61, McLachlan63a, 
Wylie85, Milonni94, Sols82, Buhmann07, Bezerra08, Bimonte09b, Skagerstam09}, 
it provides a physically transparent way to reach our final result.

\subsection{Energy of a polarizable particle in an electromagnetic field}
\label{sec:polarization-energies}

When a polarizable particle is introduced in an
electric field, the change in energy of the system is given by \cite{Jackson75}
\begin{equation}
\mathcal{F}=
- \frac{\langle\mathbf{d}(t) \cdot \mathbf{E}(\mathbf{r}_{0},t)\rangle}{2}
\label{eneq}~.
\end{equation}
where, since we are working in the Heisenberg representation, all the operators are 
time dependent. From the thermodynamic point of view the previous quantity is 
a \emph{free energy} and gives the amount of work that can be extracted from 
the system by moving the particle: in our (thermodynamic) convention a negative 
free energy will correspond to an attractive interaction (binding energy).

The expectation value $\langle \cdots\rangle$ is taken over the (initial) state 
of the non-coupled system;
% and $\langle\cdot\rangle_{T}$ denotes the thermal average.
$\mathbf{d}$ is the (electric or magnetic) dipole operator 
and $\mathbf{E}$ the corresponding
field operator, evaluated at the dipole position $\mathbf{r}_{0}$. We are implicitly 
assuming that the size of the particle is small enough to locally probe the 
electromagnetic field. 
The factor $1/2$ in \eqref{eneq} arises from the fact that we are considering the 
energy of a linear polarizable particle in an external field, rather than a 
permanent dipole \cite{Jackson75}. Note that the choice of a particular ordering 
does not seem to be necessary at this stage since the dipole operator and the 
electric field operator commute. The symmetric order proves, however, to be 
particularly useful if we want to attach a physical meaning to each single 
contribution to the energy \cite{Dalibard82,Dalibard84}, see 
Eq.(\ref{eq:F-split-in-fl}).

The Hamiltonian of the coupled system can be in general written as 
${H} = {H}_{0}+{V}$, where ${H}_{0}$ is  the sum of the Hamiltonians 
of the two isolated subsystems and ${V}$ describes the interaction between 
them. Starting from this,
the equation of motion for an operator ${A}( t )$  
%
% a generic operator in the Heisenberg picture $\mathbf{A}(t)$ 
can be written in the following integral form
(Heisenberg picture)
\begin{equation}
{A}(t) = {A}^{\rm free}(t) + 
\frac{\imath}{\hbar}\int_{0}^{t}\!{\rm d}\tau\;
e^{\frac{\imath}{\hbar}{H}_{0} (t - \tau)}
[
{V}, 
{A}(\tau)
]\;
e^{\frac{\imath}{\hbar}{H}_{0} (t - \tau)}
\label{integral-equation}~,
\end{equation}
%\begin{equation}
%\rho(t) = \rho^{\rm free}(t) - 
%e^{- \frac{\imath}{\hbar}\mathbf{H}_{0}t} 
%\frac{\imath}{\hbar}
%\int_{0}^{t}\!{\rm d}t' \,
%[\mathbf{V}^{\rm free}(t'), 
%e^{\frac{\imath}{\hbar}\mathbf{H}_{0}t'} \rho(t') 
%e^{-\frac{\imath}{\hbar}\mathbf{H}_{0}t'}]
%e^{\frac{\imath}{\hbar}\mathbf{H}_{0}t} 
%	\label{integral-equation}
%\end{equation}
%
where $t=0$ was chosen as initial time and the superscript free indicates that the 
operator evolves with respect to the Hamiltonian of the uncoupled system 
(${H}_{0}$), i.e.
\begin {equation}
{A}^{\rm free}(t) = 
e^{\frac{\imath}{\hbar}{H}_{0}t} {A} 
e^{-\frac{\imath}{\hbar}{H}_{0}t}~.
% \quad\text{and}\quad
%\rho^{\rm free}(t) =  e^{-\frac{\imath}{\hbar}\mathbf{H}_{0}t}
%\rho e^{\frac{\imath}{\hbar}\mathbf{H}_{0}t}
%
% \quad\text{and}\quad
%
% \mathbf{V}^{\rm free}(t) = e^{\frac{\imath}{\hbar}\mathbf{H}_{0}t}
% \mathbf{V} e^{-\frac{\imath}{\hbar}\mathbf{H}_{0}t}
\end{equation}
Now, within first-order perturbation theory, eq.\eqref{integral-equation} can be 
solved by replacing the 
operator ${A}(t')$ under the integral by its corresponding free evolution. 
If the interaction Hamiltonian is bilinear like in the case of the electric dipole 
interaction, ${V}= -\mathbf{d} \cdot \mathbf{E}(\mathbf{r}_{0})$, 
we get the following approximate expression for the dipole:
\begin{subequations}
\begin{align}
\mathbf{d}(t) &\approx \mathbf{d}^{\rm free}(t) + \mathbf{d}^{\rm in}(t)~,\\
\mathbf{d}^{\rm in}(t) &= 
\int_{-\infty}^{\infty}\!{\rm d}\tau
\left(\frac{\imath}{\hbar}[\mathbf{d}^{\rm free}(t),
\mathbf{d}^{\rm free}(t-\tau)] \theta(\tau)\right)
\cdot \mathbf{E}^{\rm free}(\mathbf{r}_{0}, t-\tau)
	\label{eq:dipole-commutator}~,
\end{align}
\label{dipole}
\end{subequations}
and similarly for the field:	
\begin{subequations}
	\label{eq:split-field}
\begin{align}
\mathbf{E}(\mathbf{r},t) & \approx 
\mathbf{E}^{\rm free}(\mathbf{r}, t) 
+ \mathbf{E}^{\rm in}(\mathbf{r}_{0}, t)~,\\
\mathbf{E}^{\rm in}(\mathbf{r}_{0}, t) &=
\int_{-\infty}^{\infty}\!{\rm d}\tau
\left(\frac{\imath}{\hbar}
[\mathbf{E}^{\rm free}(\mathbf{r},t),
\mathbf{E}^{\rm free}(\mathbf{r}_{0},t-\tau)] 
\theta(\tau)\right)
\cdot \mathbf{d}^{\rm free}(t-\tau)
	\label{eq:field-commutator}~.
\end{align}
The $\tau$-integral runs effectively over $\tau \ge 0$ (note the
step function $ \theta(\tau)$) because in Eq.(\ref{integral-equation}),
only times $\tau > 0$ after the initial time are relevant (causality). 
In addition, we have set the upper limit to $\tau = \infty$ assuming
that there exists a transient time $\tau_c$ after which the system
behaviour becomes stationary.
This time can be estimated
from the system operators in Eqs.(\ref{eq:dipole-commutator},
\ref{eq:field-commutator}): the commutators are either c-number functions
that die out for time arguments that differ by more than $\tau_c$,
or taking the expectation value, one gets subsystem correlation functions
with $\tau_c$ as correlation time.
%
% to preserve the \emph{causality}, i.e. the evolution of system cannot depend on 
%times preceding the initial time ($t<0$). We will be interested in equilibrium 
%(asymptotic) behaviors ($t\gg \tau_{c}$ where $\tau_{c}$ is the characteristic 
%time of the corresponding evolution) we have extended the upper boundary of the 
%integration to infinity[\texttt{add a comment on the validity of the approximation 
%following Dalibard84}]. 

We have therefore that, within the first order perturbation theory, the dipole in 
addition to its unperturbed evolution ($\mathbf{d}^{\rm free}(t)$) ``responds'' 
linearly ($\mathbf{d}^{\rm in}(t)$) to an external perturbation (in this case the 
electromagnetic field). The same also happens to the electromagnetic field where 
now the dipole is the external source of perturbation. The term in parenthesis 
under the integrals~(\ref{eq:dipole-commutator},
\ref{eq:field-commutator}), 
when evaluated over a particular state, is called \emph{susceptibility}
and contains the detailed physical information about the linear response of the 
system to the perturbation \cite{Dalibard82,Dalibard84}. In the particular case of 
a dipole the Fourier-transform of the susceptibility tensor is the polarizability
\end{subequations}
\begin{equation}
\int_{-\infty}^{\infty}\!{\rm d}t\,
\frac{\imath}{\hbar}\langle a\vert[\mathbf{d}^{\rm free}(t),\mathbf{d}^{\rm 
free}(0)] \vert a \rangle 
\theta(t)
e^{\imath \omega t} 
= \overleftrightarrow{\alpha}^{a}(\omega)~.
	\label{polarizability}
\end{equation}
where we have taken the expectation value for a given quantum state 
$| a \rangle$.
In the time domain (see Eq.(\ref{eq:dipole-commutator})), the atomic
susceptibility links the hermitean dipole operator to a hermitean field
operator; hence it must be a real function. The polarizability, being its
Fourier transform, therefore satisfies 
\begin{equation}
[\overleftrightarrow{\alpha}^{a}(\omega)]^{*} =
\overleftrightarrow{\alpha}^{a}(-\omega^{*})
	\label{symmetry}~.
\end{equation}
In addition, because of causality, Eq.(\ref{polarizability}) implies that
$\overleftrightarrow{\alpha}^{a}(\omega)$ must be analytical in 
the upper-half of the complex $\omega$-plane.

Similar conclusions hold for the electromagnetic field. If we assume that the 
dynamics of the field and the surrounding matter (other than the atom)
can be completely described in terms of bosonic operators \cite
{Huttner92,Barnett96,Buhmann07}, %, Kheirandish08,Kheirandish08a} 
the result of the commutator in \eqref{eq:field-commutator} is a c-number and the 
susceptibility does not depend on the state of the radiation. The linearity of the 
Maxwell equations then ensures that the result of eq.\eqref{eq:split-field} 
for a point-like dipole is correct to all orders. A simple identification leads to 
the following expression:
\begin{equation}
\int_{-\infty}^{\infty}\frac{\imath}{\hbar}
[
\mathbf{E}^{\rm free}(\mathbf{r},t),
\mathbf{E}^{\rm free}(\mathbf{r}_{0},0)
] \theta(t) 
e^{\imath \omega t} {\rm d}t
= \overleftrightarrow{G}(\mathbf{r}, \mathbf{r}_{0}, \omega)~.
\end{equation}
where $ \overleftrightarrow{G}$ is the electric field Green tensor, solution to 
Maxwell's equation
\begin{equation}
-\nabla_{\mathbf{r}}\times\nabla_{\mathbf{r}}\times 
\overleftrightarrow{G}(\mathbf{r}, \mathbf{r}_{0}, \omega)
+\frac{\omega^{2}}{c^{2}}\epsilon(\mathbf{r},\omega) 
\overleftrightarrow{G}(\mathbf{r}, \mathbf{r}_{0}, \omega)
= -\frac{\omega^{2}}{ \varepsilon_0 c^{2}}
\delta(\mathbf{r}-\mathbf{r}_{0}) \overleftrightarrow{I}
	\label{eq:solved-by-Green-tensor}~,
\end{equation}
where $ \overleftrightarrow{I}$ is the identity tensor and 
$\epsilon(\mathbf{r},\omega) $ is the local dielectric function (here supposed to be a scalar for simplicity) of the 
matter surrounding the dipole.
In conclusion, in frequency space the induced quantities can be described 
in terms of the retarded response functions \cite{Jackson75}
\begin{eqnarray}
&\mathbf{d}^{\mathrm{in}}(\omega) = \overleftrightarrow{\alpha}(\omega)\cdot 
\mathbf{E}^{\mathrm{free}}(\mathbf{r}_{0},\omega) 
\label{eq:def-induced-field}~,
\\
&\mathbf{E}^{\mathrm{in}}(\mathbf{r},\omega) = 
 \overleftrightarrow{G}(\mathbf{r},\mathbf{r}_{0}, \omega)
\cdot\mathbf{d}^{\mathrm{free}}(\omega)~,
\nonumber\label{eq:field-G-and-dipole}
\end{eqnarray}
where the frequency dependence and causality allow for a temporal delay.  This is slightly schematic because the polarizability tensor is defined only 
when the average is taken. 

Expressions \eqref{dipole} and \eqref{eq:split-field} formalize the considerations 
made 
at the beginning of this section:  both the dipole moment $\mathbf{d}(t)$ and 
the field $\mathbf{E}(t)$
can be split into two parts, the (free) fluctuating part describes the 
free intrinsic fluctuation, while the induced part arises in perturbation theory from
the dipole coupling \cite{Mandel95}.
Eq.\eqref{eneq} becomes
\begin{equation}
\mathcal{F}=-\frac{\langle \mathbf{d}^{\mathrm{in}}(t) 
; \mathbf{E}^{\mathrm{free}}(\mathbf{r}_{0},t)\rangle}{2}
- \frac{\langle 
\mathbf{E}^{\mathrm{in}}(\mathbf{r}_{0},t)
;
\mathbf{d}^{\mathrm{free}}(t) 
\rangle}{2}~.
	\label{eq:F-split-in-fl}
\end{equation}
We assumed a factorized initial state in which each free evolution operator is
zero on average and where the correlations between the fluctuating parts
are entirely encoded in the linear response functions, setting the correlation between the freely fluctuating components to zero.
This assumption would break down at higher orders
of perturbation theory. 
Note that while in Eq.(\ref{eneq}) the total dipole and field operators (Heisenberg 
picture) commute at equal times, this is no 
longer true for their `in' and `free' constituents in Eq.(\ref{eq:F-split-in-fl}). The 
choice of the symmetric order (indicated by the semicolon) 
allows one to see each term of the previous expression as the result of the 
quantum expectation value of a Hermitian operator and therefore to attach to it a 
physical meaning \cite{Dalibard82, Dalibard84,Meschede90}. The first term on the 
right hand side of \eqref{eq:F-split-in-fl} can be seen as the contribution to the 
Casimir-Polder energy coming from the fluctuations of the vacuum field; the second 
will be called the self-reaction term since it arises from the interaction of the 
dipole with the field generated by the dipole itself.
%although the operators involved in each single term of the previous expression do 
%not commute, the whole result does not depend on the order because $
%[\boldsymbol{\mu}(t),\boldsymbol{B}(\boldsymbol{r}_{0},t)]=0$, i.e. the total 
%dipole and field operator do commute. 

\subsection{Equilibrium fluctuations}

Consider now a configuration at global thermal equilibrium, i.e. when both the 
dipole and 
the field are in a thermal state at temperature $T$.
In this case we can apply the \emph{fluctuation-dissipation theorem} \cite
{Callen51}. This milestone of the linear response theory connects the correlation of 
a generic observable of a system in thermal equilibrium at temperature $T$ with 
the imaginary part of the linear susceptibility which characterizes  
the response to a weak perturbation.
In our case the theorem holds separately for the dipole and the field 
and we have  \cite{Agarwal75,Eckhardt82,Eckhardt84}
\begin{eqnarray}
\langle E_i^{\mathrm{free}}(\mathbf{r},\omega)
	E_j^{\mathrm{free}}(\mathbf{r},\omega') \rangle_T
&=& 2\pi\hbar\, \delta(\omega+\omega')
\coth\left(\frac{\hbar \omega}{2k_B T}\right)
\im\![G_{i j}(\mathbf{r}, \mathbf{r}, \omega)]
~,
	\label{eq:FTD_field}
\\
\langle d_i^{\mathrm{free}}(\omega)d_j^{\mathrm{free}}(\omega')\rangle_T
&=& 2\pi\hbar\, \delta(\omega+\omega')
\coth\left(\frac{\hbar \omega}{2 k_B T}\right)
\im\![ \alpha_{i j}^{T}(\omega)]~,
	\label{eq:FTD_dipole}
\end{eqnarray}
where the symbol $\langle\cdots\rangle_{T}$ define the quantum and the thermal average and, according to \eqref{polarizability}, 
$\overleftrightarrow{\alpha}^{T}(\omega)$ defines
the atomic polarizability 
operator evaluated at temperature $T$ (see Eq.(\ref
{eq:WS_polarizability_thermal}) below). 
The $\coth$ function in Eqs.(\ref{eq:FTD_field}, \ref{eq:FTD_dipole})
arises from the symmetrically ordered average of bosonic annihilation
and creation operators in a state of thermal equilibrium:
\begin{equation}
	\langle a^\dag a + a a^\dag \rangle_{T} = 
	1 + 2 N( \omega ) = \coth\Big(
	\frac{\hbar \omega}{2 k_B T}\Big)
	\label{eq:recall-origin-coth}~,
\end{equation}
where $N$ is the Bose-Einstein distribution. 
Note the asymptotic limits 
\begin{equation}
	\coth\frac{\hbar \omega}{2 k_B T} \to 
	\begin{cases}
	1 & T \ll \hbar\omega/k_{B}\\
%	\mbox{ or } \quad\to 
	\frac{2 k_B T}{\hbar \omega}& T \gg \hbar\omega/k_{B}
	\end{cases}
	\label{eq:coth-asymptotics}
\end{equation}
in the ``quantum'' (low-temperature) and ``classical'' (high-temperature) limits.

The expression given in \eqref{eq:FTD_field} can be directly reconnected with the 
currents fluctuating inside the media surrounding the dipole. For these currents 
Rytov's theory \cite{Rytov53} predicts a correlation similar to \eqref
{eq:FTD_dipole} where the role of the polarizability is now played by the dielectric 
function \cite{Agarwal75,Eckhardt82,Eckhardt84} (see 
Sec.\ref{s:non-equilibrium} below). This picture also lends itself to a natural
generalization where the bodies are assumed to be in ``local thermal equilibrium''
(see Sec.\ref{s:non-equilibrium-field}).

Note that the field correlations are needed at the same position $\mathbf{r}_0$. 
The Green function is, however, divergent in this limit due to its free-space 
contribution 
\begin{equation}
 \overleftrightarrow{G}(\mathbf{r}, \mathbf{r}_{0},\omega)= \underbrace
{\overleftrightarrow{G}_{0}(\mathbf{r}, \mathbf{r}_{0},\omega)}_{\text{free 
space}}+ \underbrace{\overleftrightarrow{\mathcal{G}}(\mathbf{r}, \mathbf{r}_
{0},\omega)}_{\text{scattered}}~.
	\label{splitting-G}
\end{equation}
The corresponding part of the free energy provides the Lamb shift of the internal 
levels of the dipole immersed in the electromagnetic field \cite{Milonni94}. This 
contribution is position-independent and does not contain any information about 
the interaction between the bodies and the dipole. Therefore it can be safely 
``hidden'' in the (renormalized) energy levels of the atom. 
The physical information about the interaction is indeed contained only in the 
\emph{scattered} part of the Green function \cite{Wylie85}.
If the body happens to be a plane surface, it follows from symmetry
that the result can only depend on the dipole-surface distance $L$ and we can set 
$\overleftrightarrow{\mathcal{G}}(\mathbf{r}_{0};\mathbf{r}_{0},\omega)\equiv
\overleftrightarrow{\mathcal{G}}(L,\omega)$. 

Combining Eqs.(\ref{eq:F-split-in-fl}--\ref{eq:FTD_dipole}), we  finally obtain that
the free energy of a polarizable
particle at nonzero temperature $T$ has the following general form
 (Einstein summation convention)
%%%%%%%%%%%%%%%%%
\begin{equation}
\mathcal{F}= -\frac{\hbar}{2\pi}
%{\rm Tr}
\int\limits_{0}^{\infty}{\rm d}\omega
\coth\!\left( \frac{\hbar\omega}{2k_{B}T} \right)
%\im\![\boldsymbol{\beta}^{T}(\omega) \cdot 
% \boldsymbol{\mathcal{G}}(L, \omega)]
\im\![\alpha^{T}_{ij}(\omega) \mathcal{G}_{ji}(L, \omega)]
\label{eq1:result-CP-free-energy}~,
\end{equation}
We have used the reality condition~\eqref{symmetry}, implying that 
the imaginary part of both polarizability and Green tensors are odd 
in $\omega$. Eq.\eqref{eq1:result-CP-free-energy} coincides with the expression of 
the  atom-surface interaction derived by many authors \cite
{Casimir48a,Dzyaloshinskii61, McLachlan63a, Wylie85, Milonni94, Sols82, 
Buhmann07, Bezerra08, Bimonte09b, Skagerstam09, 
Meschede90,Haakh09b,Henkel02,Novotny08}.  It is often expressed in an 
equivalent form using the analyticity of
$\overleftrightarrow{\alpha}^{T}(\omega)$ 
and $\overleftrightarrow{\mathcal{G}}(L, \omega)$ in the upper half of
the complex frequency plane.
Performing a Wick rotation in the complex frequency plane yields the so-called 
Matsubara expansion \cite{Matsubara55,Dzyaloshinskii61}
\begin{equation}
\label{eq:Matsubara-series}
\mathcal{F}(L, T)=-k_{B}T
%{\rm Tr}
\msum
\alpha^{T}_{i j}(\imath \xi_n)
%\cdot
\mathcal{G}_{j i}
(L, \imath \xi_n)~,
\end{equation}
where the Matsubara frequencies 
$\xi_{n} = 2\pi n k_{B}T/\hbar$ 
arise from the poles of the hyperbolic cotangent,
%$\coth( \hbar \omega / 2 k_B T )$,
and the prime in the sum indicates that the $n=0$ term comes with a 
coefficient $1/2$.
Both $\overleftrightarrow{\alpha}^{T}(\imath \xi)$ and $\overleftrightarrow
{\mathcal{G}}(L, \imath \xi)$ are real expressions for $\xi>0$ 
because of Eq.(\ref{polarizability}).

These considerations conclude our first general analysis of the Casimir-Polder 
interaction. In the following section we will analyze the particle response
function appearing 
in the previous formulation, namely the atomic polarizability, and mention
also the case of a nanoparticle.

%==================================================
\subsection{Polarizability tensor}
\label{s:beta-and-alpha}
%==================================================

The previous results can be used for the interaction of a surface with atoms, 
molecules, particles or in general any (small) object that can be described with 
good approximation in terms of a electric-dipole
polarizability tensor. Here we are going to review 
the polarizability of an atom and of a  nanoparticle.

\subsubsection{Atoms}

The polarizability tensor is determined by the transition dipole matrix elements
and the resonance frequencies. 
For an arbitrary atomic state $|a\rangle$ it can be written as
\begin{equation}
\alpha_{ij}^a({\omega})=\sum_b
\frac{d_i^{ab}d_j^{ba}}{\hbar}\frac{2\omega_{ba}}{\omega_{ba}^{2}-(\omega+
\imath 0^{+})^{2}}
\label{eq:WS_polarizability}~,
\end{equation}
%Eq.(\ref{eq:WS_polarizability}) above.
%(For the comparison to a small metallic sphere, we give
%the magnetic polarizability in Appendix~\ref{a:para-vs-dia}.)
where $d_i^{ba}$ is the matrix element between the states $|b\rangle$ and $|a
\rangle$ of the $i$ component of the electric
dipole operator and $\omega_{ba} = (E_{b} - E_{a}) / \hbar$
the 
corresponding transition frequency. The introduction of an infinitesimal imaginary 
part shifts the poles of the expression into the lower part of the complex frequency 
plane ($\pm \omega_{ba}-\imath 0^{+}$), which is mathematically equivalent to 
the causality requirement. The tensorial form of the previous expression allows to 
take into account a possible anisotropic response of the atom to an electric field. 
A simplification of the previous expression can be obtained averaging over the 
different levels and directions so that the polarizability tensor becomes
$\alpha_{ij}^{a} = \delta_{ij} \alpha^{a}_{\rm iso}$ with the scalar function
\begin{equation}
\alpha_{\rm iso}^a({\omega}) = 
\sum_b
\frac{\vert {\bf d}^{ba}\vert^{2}}{3\hbar}\frac{2\omega_{ba}}
{\omega_{ba}^{2}-(\omega+\imath 0^{+})^{2}}
\label{isotropic}~.
\end{equation}
The polarizability is exactly isotropic
when several excited sublevels that are
degenerate in energy are summed over, like the $np_{x,y,z}$ orbitals of
the hydrogen-like series.
When the atom is in thermal equilibrium, we have to sum the polarizability
over the states $|a\rangle$ with a Boltzmann weight:
\begin{equation}
\alpha^{T}_{ij}(\omega) = \sum_{a}
\frac{e^{- E_{a} / k_B T}}{Z}
	\alpha^{a}_{ij}(\omega)
\label{eq:WS_polarizability_thermal}~,
\end{equation}
where $Z$ is the partition function. In the limit $T\rightarrow 0$, we recover the 
polarizability for a ground state atom. For a single pair of levels $|a\rangle$
and $|b\rangle$, this leads to the following relation between the 
state-specific and the thermalized polarizabities:
\begin{equation}
	\alpha_{ij}^{T}( \omega ) \approx 
	\alpha_{ij}^{a}( \omega )
	\tanh\frac{ \hbar \omega_{ba} }{ 2 k_B T }
	\label{eq:thermalized-polarizability}~.
\end{equation}
This is mainly meant to illustrate the temperature dependence, otherwise it
is a quite crude approximation. The reason is that the coupling to other levels
makes the polarizabilities $\alpha_{ij}^{a}$ and $\alpha_{ij}^{b}$ differ quite a lot.
Electronically excited states are much more polarizable due to their larger electron
orbitals.

% %($|a\rangle=|g\rangle$)
%as calculated by Wylie \& Sipe \cite{Wylie85} .
%
%For a two-level system with transition frequency $\Omega_e$, 
%the previous expression takes a simple form and can be expressed 
%in terms of the ground state polarizability [Eq. (\ref{eq:WS_polarizability}), 
% where $a=g$]:
%%
%\begin{equation}\label{eq:two-level_polarizability}
%\overleftrightarrow{\alpha}^{T}(\omega) 
%= 
%\tanh\left(\frac{\hbar\Omega_e}{2k_B T}\right) 
% \overleftrightarrow{\alpha}^{g}(\omega)
%~.
%\end{equation}
%%

\subsubsection{Nanospheres}

% We have seen now, that the polarizability of an atom takes a positive constant 
% value at low frequency and the induced magnetic dipole is parallel to the magnetic 
% field (paramagnetism).
Let us consider now the case where the atom is replaced by a nanosphere \cite
{Novotny08,Haakanson08,Klimov09}. Indeed, if the sphere radius $R$ is smaller 
than the penetration depth and the radiation wavelength, we can neglect higher 
order multipoles in the Mie expansion \cite{Mie08} and consider only the electric 
and magnetic dipole (the sphere is globally neutral). 

In this long-wavelength limit,
the Clausius-Mossotti relation \cite{Jackson75,Born99} provides 
the electric polarizability 
\begin{equation}
\alpha_{\rm sph}(\omega) = 
4\pi \epsilon_{0} R^{3}\frac{\epsilon(\omega)-1}{\epsilon(\omega)+2}
	\label{eq:def-alpha-nanosphere}~.
\end{equation}
where $\epsilon( \omega )$ is the (scalar) dielectric function of the 
sphere material.
The nanosphere has also a magnetic polarizability that
arises because a time-dependent magnetic field induces circulating
currents (Foucault currents) 
\cite{Jackson75}. This leads to a diamagnetic response
\cite{Feinberg70} 
\begin{equation}
	\label{eq:def-beta-nanosphere}
\beta_{\rm sph}(\omega) = \frac{2\pi}{15 \mu_0}
\left(\frac{R\omega}{c}\right)^{2}[\varepsilon(\omega) - 1]R^{3} 
~.
\end{equation}
Both polarizabilities are scalars.  For a metallic sphere, the
electric polarizability goes to a positive constant 
at zero frequency, while the magnetic 
one vanishes there and has a negative real part 
at low frequencies (diamagnetism).

For a qualitative comparison to an atom, one can estimate the oscillator strength \cite{Mandel95}, 
defined by the integral over the imaginary part of the polarizability. 
For the atom we have
\begin{equation}
\int_0^\infty \!{\rm d}\omega \, {\rm Im}\; \alpha_{\rm at}(\omega) 
\sim \frac{\pi (e a_{0})^2}{\hbar}
\quad\text{and}\quad
\int_0^\infty\! {\rm d}\omega \, {\rm Im}\; \beta_{\rm at}(\omega) 
\sim \frac{\pi \mu_B^2}{\hbar}~,
\end{equation} 
where the Bohr radius $a_0$ and the Bohr magneton $\mu_B$ give the
overall scaling of the transition dipole moments. The following dimensionless
ratio allows a comparison between the two:
\begin{equation}
	\frac{(e a_0)^2 / \varepsilon_0 
	}{ 
	\mu_B^2 \mu_0 }
	\sim \frac{1}{\alpha^{2}_{\rm fs}}
	\label{eq:compare-oscillator-strengths}~,
\end{equation}
where $\alpha_{\rm fs} = e^2 / (4\pi\varepsilon_0 \hbar c)
\approx 1/137$ is the fine structure constant. The
electric oscillator strength clearly dominates in the atom.

Let us compare to a metallic nanosphere (gold is often used in experiments) 
and assume a Drude model~\eqref{DrudeModel} for the dielectric function.
In terms of the volume $V=4\pi R^{3}/3$, we get an electric oscillator
strength
\begin{eqnarray}
\int_0^\infty {\rm Im}\; \alpha_{\rm sph}(\omega) d\omega &=&\frac{3}{2}
\epsilon_0  \frac{\omega_{ p}}{\sqrt{3} } V + \mathcal{O}(\frac{\gamma}
{\omega_p})~.
\end{eqnarray}
where $\omega_{p}/\sqrt{3}$ is the resonance frequency of the particle
plasmon mode (the pole of $\alpha_{\rm sph}( \omega )$, 
Eq.(\ref{eq:def-alpha-nanosphere})). This is much larger than for an atom
if the nanoparticle radius satisfies $a_0 \ll R \ll \lambda_p$, 
i.e., a few nanometers.
The magnetic oscillator strength can be estimated as
\begin{equation}
\int_0^{\omega_p} {\rm Im}\; \beta_{\rm sph}(\omega) d\omega = \frac{2\pi}
{3 \mu_0}  \gamma \log(\frac{\omega_p}{\gamma})
\left(\frac{R}{\lambda_{p}}\right)^2 V~.
\end{equation}
where we took $\omega_p$ as a cutoff frequency to make the integral 
convergent (at higher frequencies, Eq.(\ref{eq:def-beta-nanosphere}) does
not apply any more). 
We have used the plasma wavelength $\lambda_{p}=2\pi c/\omega_{p}$ ($\sim 
100$ nm for gold). 
Similar to an atom, the nanoparticle response is dominantly electric, 
but the ratio of oscillator strengths can be tuned via the material parameters 
and the sphere size. The magnetic contribution to the particle-surface interaction
is interesting because it features a quite different temperature dependence,
see Ref.\cite{Haakh09b}.

\subsection{Non-perturbative level shift}
	\label{sec:non-perturbative-level-shift}
	
In the previous section we saw that the main ingredient to derive the Casimir-Polder 
interaction between a particle and an object is the ability to solve for the dynamics 
of the joint system particle+electromagnetic field. Previously we limited ourself to a 
solution at the first order in the perturbation, implicitly motivated by the difficulty 
to solve \emph{exactly} the dynamics of a multi-level atomic system 
coupled to a continuum of bosonic degrees of freedom (e.m.\ field). Things are 
different if we consider the linear coupling between two bosonic systems, i.e. if we 
describe the particle as a quantum harmonic oscillator. The linearity of the coupled 
system allows for an exact solution of its dynamics and even if the 
harmonic oscillator may be in some cases only a poor description of an atom \cite
{Babb05}, it is a good representation of a nanoparticle (the resonance
frequency being the particle plasmon frequency). Generally, this approach
gives a first qualitative 
indication for the physics of the interaction\cite
{Renne71a,Renne71,Davies72,Mahanty72,Mahanty73}.

The main idea we follow in this section is based upon a generalization of the
``remarkable formula'' of Ford, Lewis and O'Connell \cite{Ford85,Ford88}  
(see 
also \cite{Renne71a,Renne71,Davies72,Mahanty72,Mahanty73}). According 
to this formula, 
the free energy of a one-dimensional oscillator immersed in black body 
radiation is
\begin{equation}
\mathcal{F}_{\rm FLOC}(T) = \frac{1}{\pi}\int_{0}^{\infty}{\rm d}\omega\ f(\omega, T){\rm 
Im}\left[\partial_{\omega}\ln \alpha(\omega)\right]
	\label{eq:remarkable-FLOC}~,
\end{equation}
where $f(\omega, T)$ is the free energy per mode, 
\begin{equation}
f(\omega, T)=k_{B} T\log\left[2 \sinh\left(\frac{\hbar\omega}{2k_{B}T}\right)
\right]~,
\end{equation}
and $\alpha(\omega)$ is the (generalized) susceptibility of the oscillator
derived from Eq.(\ref{oscillator2}) below. More precisely,
$\mathcal{F}(T)$ gives the difference between two free energies: the
oscillator coupled to the radiation field and in equlibrium with it, on the
one hand, and solely the radiation field, on the other. 
Eq.(\ref{eq:remarkable-FLOC}) is ``remarkable'' because the only 
system-relevant information needed here is the susceptibility function.

In three dimensions, the polarizability becomes a tensor
\begin{equation}
\label{eq:linearResponse}
\mathbf{d}(\omega) = \overleftrightarrow{\alpha}(\omega)\cdot\mathbf{E}
(\omega)~,
\end{equation}
where $\mathbf{E}(\omega)$ is the external electric field. In the case 
considered by Ford and O'Connell, there was no need to include a spatial
dependence because of the homogeneity and isotropy of the black body 
field. We are going to consider this symmetry to be broken by the presence 
of some scattering object. 
As a consequence, the generalized susceptibility tensor becomes position-%
dependent $\overleftrightarrow{\alpha}(\omega, {\bf r}_0)$. The spatial dependence is connected with the scattered part of the Green function and 
leads both to a position-dependent frequency renormalization and a damping rate.

In order to get the expression of $\alpha(\omega, {\bf r}_0)$, let us consider 
for simplicity 
the equation of motion of an isotropic oscillator with charge $q$ interacting with 
the e.m. field near some scattering body (that is described by a dielectric 
constant). In frequency space, the (nonrelativistic) dynamics of the oscillator 
is described by
%\begin{equation}
%\mathbf{H}=\frac{1}{2m}\left[\boldsymbol{\pi}^{2}+\omega_{0}^{2}\, 
% \mathbf{x}^{2}\right]+q\phi(\mathbf{r_{0}})+\mathbf{H}_{R}
%\end{equation}
%with
%\begin{equation}
%\boldsymbol{\pi}=\left(\mathbf{p}-\frac{q}{c}\mathbf{A(r_{0})}\right)
%\end{equation}
%where $\mathbf{H}_{R}$ is the hamiltonian of the free (vacuum) field and 
% $\mathbf{A(r_{0})}$ and $\phi(\mathbf{r_{0}})$ are the vector and the scalar 
% potential, respectively\footnote{The e.m. field is evaluated on the position of the 
% oscillator and we have also assumed that the electromagnetic field is spatially 
% constant over the its whole dimension}.  
% Starting from there we can derive the following equations of motion 
\begin{equation}
m\left[\omega^{2} \mathbf{d}(\omega) + 
\omega^{2}_{0} \mathbf{d}(\omega)
\right] = q^2 \mathbf{E}(\mathbf{r}_{0}, \omega)
%+\frac{q}{2c}\left[\dot{\mathbf{x}}(t)\times \mathbf{B}(t,\mathbf{r}_{0})-
% \mathbf{B}(t,\mathbf{r}_{0})\times \dot{\mathbf{x}}(t)\right]
\label{oscillator}~,
\end{equation}
where we have neglected the coupling with the magnetic field (first order in 
$\dot d/c$). For the field we have
\begin{gather}
%\nabla\times \mathbf{B}(t,\mathbf{r}) -
% \frac{1}{c}\dot{\mathbf{D}}(t,\mathbf{r})=\frac{4\pi}{c}\mathbf{j}(t)\\
\nabla\times\nabla\times \mathbf{E}(\mathbf{r}, \omega) - 
\frac{\omega^{2}}{c^{2}}
\epsilon(\omega, \mathbf{r})\mathbf{E}(\mathbf{r}, \omega)
= \imath\omega \mu_0 \mathbf{j}(\mathbf{r}, \omega)~,
\end{gather}
%\begin{gather}
%\dot{\mathbf{x}}=\frac{\boldsymbol{\pi}}{m}\\
%\dot{\boldsymbol{\pi}}=-\omega^{2}_{0}\frac{\mathbf{x}}{m}-
% q\nabla\phi (\mathbf{r_{0}})
%\end{gather}
%\begin{gather}
%\label{oscillator}
%-m(\omega^{2}-\omega_{0}^{2})\mathbf{x}(\omega)-\imath \omega 
% \frac{q}{c}\mathbf{A}(\omega,r_{0})=0
%\end{gather}
where the source current is 
$\mathbf{j}(\mathbf{r}, \omega)=-\imath
\omega \mathbf{d}(\omega)\delta(\mathbf{r} - \mathbf{r}_{0})$. \\

Now, the formally exact solution for the operator $\mathbf{E}$ can be given 
in term 
of the (electric) Green tensor:
\begin{equation}
\mathbf{E}(\mathbf{r}, \omega) = 
\mathbf{E}^{\rm free}(\mathbf{r}, \omega)
+ \overleftrightarrow{G}(\mathbf{r}, \mathbf{r}_0, \omega) 
\cdot \mathbf{d}(\omega)~,
\label{field}
\end{equation}
where the Green tensor is the solution of Eq.(\ref{eq:solved-by-Green-tensor})
given above.
%\begin{equation}
%\nabla\times\nabla\times\overleftrightarrow{G}(\omega,r;r_{0})-\frac{\omega^
%{2}}{c^{2}}\epsilon(\omega, \mathbf{r})\overleftrightarrow{G}(\omega,r;r_{0})=
%\omega^{2}\frac{4\pi}{c}\delta(r-r_{0})\overleftrightarrow{I}.
%\end{equation}
The field $\mathbf{E}^{\rm free}(\mathbf{r}, \omega)$ is the electromagnetic field we 
would have without the oscillator and it is connected with the intrinsic fluctuations 
of the polarization field, or equivalently, of the currents in the body.
Physically Eq. \eqref{field} states that the total electromagnetic field is given by 
the field present near the scattering object plus the field generated by the 
dipole. 
Introducing Eq.\eqref{field} in Eq.\eqref{oscillator} we get
\begin{gather}
\label{oscillator2}
-m(\omega^{2} - \omega_{0}^{2}) \mathbf{d}(\omega) - 
q^{2} \overleftrightarrow{G}(\mathbf{r}_{0}; \mathbf{r}_{0}, \omega) 
\cdot \mathbf{d}(\omega) 
= q^2 \mathbf{E}^{\rm free}(\mathbf{r}_{0}, \omega)~.
\end{gather}
The Green function $\overleftrightarrow{G}(\mathbf{r}; \mathbf{r}_{0}, 
\omega)$
solves an electromagnetic scattering problem and 
therefore, it decomposes naturally into a free-space field
$\overleftrightarrow{G}_0$ (as if the source
dipole were isolated in vacuum), and the field scattered by the body,
$\overleftrightarrow{\mathcal{G}}$. This is at 
the basis of the splitting in Eq.(\ref{splitting-G}) discussed above.
% into the field generated by the dipole as if it were isolated in 
%vacuum, 
%and the field reflected by the body. This is at the basis of the splitting
%\begin{equation}
% \overleftrightarrow{G}(\omega,\mathbf{r};\mathbf{r}_{0}) = 
% \overleftrightarrow{G}_{0}(\omega,\mathbf{r};\mathbf{r}_{0})
% + \overleftrightarrow{\mathcal{G}}(\omega,L)
%\end{equation}
%discussed in the previous section. 
%in the coincidence limit ($\mathbf{r}\to \mathbf{r}_{0}$) the Green tensor 
%%$\overleftrightarrow{G}(\omega,r_{0};r_{0})$ 
%is actually divergent because of its free vacuum (bulk) contribution 
% $\overleftrightarrow{G}_{0}(\omega,\boldsymbol{r};\boldsymbol{r}_{0})$. 
% Again, 
% we define the \emph{scattered} part $\overleftrightarrow{\mathcal{G}}$ of the 
% Green function as it follow
The free-space part $\overleftrightarrow{G}_0( \mathbf{r}; \mathbf{r}_0, 
\omega )$
is a scalar in the coincidence limit because of the isotropy of 
space: part of the divergence (${\rm Re}[G_{0}]$)
can be reabsorbed into mass renormalization,
$m \omega_0^2 \mapsto m \tilde \omega_0^2$,
and part (${\rm Im}[G_{0}]$) gives rise to dissipation (damping
rate $\gamma( \omega )$). Therefore Eq.\eqref{oscillator2} can be rewritten 
as
%
%which is position independent and only in a particular limit (cutoff frequency 
% charaterizing the response of the bath going to infinity) frequency independent. 
% and splitting it in real and imaginary part we get
\begin{gather}
\label{oscillator3}
\left(-\omega^{2} - \imath\gamma(\omega)\omega + 
\tilde\omega_{0}^{2} - \frac{q^{2}}{m}
\overleftrightarrow{\mathcal{G}}(\mathbf{r}_0; \mathbf{r}_0, \omega)
\right) \cdot 
\mathbf{d}(\omega) 
= \frac{q^{2}}{m} \mathbf{E}^{\rm free}(\mathbf{r}_0, \omega)~.
\end{gather}
The free electromagnetic field plays here the role of an external force 
and therefore the generalized (or ``dressed'') polarizability tensor is given 
by 
\begin{align}
\overleftrightarrow{\alpha}(\omega,L) 
% &= 
% \frac{q^{2}}{m}\left(-\omega^{2}-\imath
% \gamma(\omega)+\tilde\omega_{0}^{2}-\frac{q^{2}}{m} 
% \overleftrightarrow{\mathcal{G}}(\omega,L)\right)^{-1}
% \nonumber
%\\
	&=\alpha_{v}(\omega)\left( 1 - \alpha_{v}(\omega)
	\overleftrightarrow{\mathcal {G}}(\mathbf{r}_0; \mathbf{r}_0, \omega) \right)^{-1}
	\label{eq:result-dressed-polarizability}~,
\end{align}
where we have defined
\begin{equation}
\alpha_{v}(\omega) = \frac{q^{2}}{m}
\left(- \omega^{2} - \imath\gamma(\omega) 
+ \tilde\omega_{0}^{2}\right)^{-1}\ ~.
\end{equation}
%where for symmetry reasons we have anticipated that the reflected Green 
% function depend on the distance of the atom from the surface. 
%Moreover since $\mathbf{G}_{r}$ is diagonal also $\alpha(\omega,d)$ is 
% diagonal. 
If Ford, Lewis and O'Connell's result is generalized to a three-dimensional oscillator, 
a trace operation appears before the logarithm in Eq. \eqref{eq:remarkable-FLOC}.
Using the identity ${\rm tr}\,\log \overleftrightarrow{a} = \log\, \det  \overleftrightarrow{a}$,
one gets
\begin{align}
\label{FordGen}
\mathcal{F}_{\rm FLOC}(T)&=\frac{1}{\pi}\int_{0}^{\infty}{\rm d}\omega\ 
f(\omega, T)
{\rm Im}\left[\partial_{\omega}
\ln \det \overleftrightarrow{\alpha} (\omega, d)\right]
\nonumber\\
&=\frac{1}{\pi}\int_{0}^{\infty}{\rm d}\omega\ f(\omega, T){\rm Im}
\left[\partial_{\omega}\ln \alpha_{v}(\omega)\right]\nonumber\\
&-\frac{1}{\pi}\int_{0}^{\infty}{\rm d}\omega\ f(\omega, T){\rm Im}
\left[\partial_{\omega}\ln\det \left(1-\alpha_{v}(\omega)\overleftrightarrow{\mathcal
{G}}(\mathbf{r}_0; \mathbf{r}_0, \omega)\right)\right]~.
\end{align}
%where we have set
%\begin{gather}
%\mathbf{G}_{e}(\omega,d)= \frac{\imath \omega}{c} 
% \mathbf{G}_{r}(\omega,d)
%\end{gather}
%i.e. the free vacuum polarizability and the electric Green function [TO check].
The first term is distance-independent and 
coincides with the free energy of an isolated oscillator in the electromagnetic 
vacuum. 
It can be interpreted as a free-space Lamb shift.
The second part of Eq.
\eqref{FordGen} is distance-dependent and therefore gives rise to the 
Casimir-Polder 
interaction. With help of a partial integration, we finally get
\begin{equation}
\mathcal{F}
%_{\rm CP}
=\frac{\hbar}{2\pi}\int_{0}^{\infty}{\rm d}\omega 
\coth\left( \frac{\hbar\omega}{2k_{B}T} \right)
{\rm Im}\left[
	\ln\det\left(1 - \alpha_{v}(\omega)
	\overleftrightarrow{\mathcal{G}}(L, \omega)\right)
	\right]~.
\label{NP}
\end{equation}
The previous result can be easily generalized to the case of  an anisotropic 
oscillator by just replacing the vacuum polarizability with the respective tensor. 

The 
usual expression~(\ref{eq1:result-CP-free-energy})
for the Casimir Polder free energy
is recovered by assuming a weak atom-field interaction.
Expanding the 
logarithm to first order we get
\begin{equation}
\mathcal{F}
%_{\rm CP}
=-\frac{\hbar}{2\pi}{\rm Tr}\int_{0}^{\infty}{\rm d}
\omega \coth\left( \frac{\hbar\omega}{2k_{B}T} \right){\rm Im}\left[\alpha_{v}
(\omega)\overleftrightarrow{\mathcal{G}}(L, \omega)\right]~.
\label{eq1NP}
\end{equation}
From a scattering point of view, this approximation is equivalent to neglecting the 
multiple reflections of the electromagnetic field between oscillator and surface. 
At short distance to the surface, these reflections become relevant; the
next-order correction to the van der Waals interaction arising from~(\ref{NP}) 
is discussed in Sec.3.4 of chapter by DeKieviet {\it et al.} in this volume.

Note that although very similar, Eqs.\eqref{eq1:result-CP-free-energy} 
and \eqref{eq1NP} are not identical. Eq.\eqref{eq1:result-CP-free-energy},
applied to an oscillator atom, would have 
featured the bare polarizability
\begin{equation}
	\alpha(\omega)=\frac{q^{2}/ m}{
	\omega_{0}^{2} - (\omega + {\rm i} 0^{+})^{2}}~,
	\label{eq:bare-polarizability}
\end{equation}
where the infinitesimal imaginary part ${\rm i}0^{+}$ ensures causality.
Eq.\eqref{eq1NP} involves, on the contrary, the renormalized or 
vacuum-dressed polarizability which is causal by default. In other words,
it contains a summation over an infinite subclass of terms in the 
perturbation series.

Finally, as a general remark and in connection with the scattering interpretation of 
dispersion forces (see the chapters by Lambrecht \emph{et al.}\  
and by Rahi \emph{et al.}\ in this volume for detailed discussions on the calculation of the Casimir effect within the framework of the scattering theory), 
within the theory of two linearly 
coupled linear systems, the susceptibilities involved in the description of the 
equilibrium Casimir-Polder interaction are the \emph{isolated and dressed} ones 
(isolated scatters). This means that, within a linear response theory, or equivalently 
up to the first order in the perturbation theory, the susceptibilities are not modified 
by the presence of the other scatters but only dressed by the electromagnetic 
field. In our case, this means that $\gamma(\omega)$ or $\tilde\omega$ in 
Eqs.\eqref{NP} or \eqref{eq1NP} do not depend on $\mathbf{r}_0$. 
 
%\todo{self-consistent polarizability with $L$-dependent resonance
%frequency}

%==================================================
\section{Atoms and a planar surface}
	\label{s:Green-Tensor-planar-surface}
%==================================================

Let us consider for definiteness the Casimir-Polder potential near a planar
surface, with a distance $L$ between the atom and surface. The Green
function is in this case explicitly known and is given in the following subsection.
%Sec.\ref{s:planar-Green-function}.

\subsection{Behaviour of the Green function}
	\label{s:planar-Green-function}
	
We sketch here the qualitative behaviour of the electromagnetic
Green function near a planar surface that 
can be calculated analytically.
Let the atom (source dipole) be on the positive $z$-axis at a 
distance $L$ from a medium that occupies the half-space below 
the $xy$-plane.

%==================================================
\subsubsection{Reflection coefficients and material response}
\label{Materials}
%==================================================

The electric Green tensor 
$\overleftrightarrow{\mathcal{G}}(\mathbf{r}, \mathbf{r}_0, \omega )$ 
is needed for coincident positions ${\bf r} = {\bf r}_0$; by symmetry 
it
is diagonal and invariant under rotations in the $xy$-plane \cite
{Agarwal75a,Agarwal75,Wylie84,Wylie85}:
\begin{multline}
%\begin{split}
\overleftrightarrow{\mathcal{G}}(L,\omega) = \frac{1}{8 \pi\epsilon_{0}}
\int\limits_0^\infty k dk \, \kappa
\left[\left(r^{\rm TM}(\omega, k)+
 \frac{\omega^2}{c^2 \kappa^{2}}r^{\rm TE}(\omega, k)\right)[\mathbf{\hat x
\hat x}+ \mathbf{\hat y\hat y}]\right.\\
\left. + 2\frac{k^{2}}{\kappa^{2}}r^{\rm TM}(\omega, k) \mathbf{\hat z\hat z}
\right]e^{-2 \kappa L }~,
\label{eq:surface-Green-function}
%\end{split}
\end{multline}
where $\epsilon_{0}$ is the vacuum permittivity, $k=|\mathbf{k}|$ is the modulus 
of the in-plane wave vector. 
and $\mathbf{\hat x\hat x}$, $\mathbf{\hat y\hat y}$, $\mathbf{\hat z\hat z}$ 
are the cartesian dyadic products. 
We consider here a local and isotropic medium, excluding 
the regime of the anomalous skin effect \cite{Sondheimer52}.
The Fresnel formulae then give
the following reflection coefficients in the TE- and TM-polarization 
(also known as s- and p-polarization) \cite{Jackson75}:
\begin{equation}
r^{\rm TE}(\omega, k)=\frac{
\mu(\omega)\kappa-\kappa_{m}}{\mu(\omega)\kappa+\kappa_{m}}~,
\quad
r^{\rm TM}(\omega, k)=\frac{\epsilon(\omega)\kappa-\kappa_{m}}{\epsilon
(\omega)\kappa+\kappa_{m}}~,
\label{eq:fresnel}
\end{equation}
where $\kappa$, $\kappa_m$ are the propagation constants in vacuum
and in the medium, respectively:
\begin{equation}
\label{eq:kappa}
\kappa=\sqrt{k^2-\frac{\omega^2}{c^2}}~,\quad 
\kappa_{m}=\sqrt{k^2-\epsilon(\omega)\mu(\omega)\frac{\omega^2}{c^2}}~.
\end{equation}
The square roots are defined so that $\im{\kappa}, \im{\kappa_{m}}\le 0$ and  $
\re{\kappa}, \re{\kappa_{m}} \ge 0$. In particular $\kappa$ is either real or pure 
imaginary. The corresponding frequencies and wave vectors define two regions in 
the $(\omega, k)$ plane \cite{Born99}:
\emph{Evanescent region}  $\omega<c k$: the electromagnetic field propagates 
only parallel to the interface and decays exponentially ($\kappa > 0$) in the 
orthogonal direction. 
\emph{Propagating region} $\omega> c k$: the electromagnetic field also
propagates ($\re{\kappa}=0$) in the orthogonal direction.
Note that the magnetic Green tensor 
$\overleftrightarrow{\mathcal{H}}$ 
can be obtained from the electric one 
by swapping the reflection coefficients \cite{Henkel99}:
\begin{equation}
\label{eq:electric_GF}
\varepsilon_0
\overleftrightarrow{\mathcal{G}}\equiv \frac{ 1 }{ \mu_0 }
\overleftrightarrow{\mathcal{H}}(r^{\rm TE} \leftrightarrow r^{\rm TM})~.
\end{equation}

All information about the optical properties of the surface is encoded 
in the response functions $\varepsilon(\omega)$ 
and $\mu(\omega)$. 
% Thus, in order to 
% draw any conclusion, one has to assume a particular model for these 
% quantities. 
For the
sake of simplicity, we focus in the following on a nonmagnetic, metallic  
medium ($\mu(\omega) = 1$)
and use the Drude model \cite
{Jackson75}:
\begin{equation}
\varepsilon(\omega)=1-\frac{\omega^{2}_{p}}{\omega(\omega+\imath
\gamma)}~,
\label{DrudeModel}
\end{equation}
where $\omega_{p}$ is the plasma frequency (usually for metals in the UV regime). The dissipation 
rate $\gamma$ takes account  of all dissipative phenomena (impurities, electron-
phonon scattering, etc.) in the metal \cite{Kittel96} and generally $\gamma/
\omega_{p}\ll1$ ($\sim 10^{-3}$ for gold).

%, which includes Ohmic dissipation in a very characteristic way. 

%==================================================
\subsubsection{Distance dependence of the Green tensor}
\label{s:distance-regimes}
%==================================================
The Drude model includes Ohmic dissipation in a very characteristic way, through 
the parameter $\gamma$. This affects the physical length scales of the 
system
(see Ref.\cite{Henkel05a} for a review). In our case the relevant ones are the  
photon wavelength in vacuum $\lambda_{\omega}$ and
the skin depth in the medium $\delta_\omega$. While the first is simply given by 
\begin{equation}
\lambda_\omega = \frac{ 2\pi c }{ \omega }~,
\end{equation}
the second is defined in terms of the low frequency behavior of the dielectric 
function
\begin{equation}
	\frac{ 1 }{ \delta_\omega } = 
	\frac{ \omega }{ c }	\, {\rm Im}\sqrt{ \varepsilon( \omega ) }
%	\sqrt{\frac{2D}{\omega}}
	\approx
	\sqrt{\frac{\omega}{2D}} \quad \mbox{(for $\omega \ll \gamma$)}~,
	\label{eq:def-skin-depth}
\end{equation}
where $D= \gamma c^{2}/\omega^{2}_{p}$ is the diffusion coefficient 
for the magnetic field in a medium with Ohmic damping
\cite{Jackson75}. The skin depth 
gives a measure of the penetration of the electromagnetic field in the medium 
($
\sim 0.79\mu{\rm m}$ at 10 GHz for gold).
%\begin{equation}
%	\lambda_\omega = \frac{ 2\pi c }{ \omega }
%	.
%	\label{eq:def-photon-wavelength}
%\end{equation}
%Note that $\varepsilon( \omega ) \approx 2{\rm i}\lambda_\omega^2/(2\pi\delta_\omega)^2$
%for frequencies $\omega \ll \gamma \ll \omega_p$ (Hagen--Rubens regime).
%This is the relevant regime for the relatively low magnetic resonance 
%frequencies \cite{Haakh09b}. 
If we have $\delta_\omega \ll \lambda_\omega$,
the dependence of the Green function on $L$
is quite different in the following three domains:
(i) the \textit{sub-skin-depth region}, $L\ll \delta_\omega$,
(ii) the 
\textit{non-retarded region}, $ \delta_\omega \ll L\ll \lambda_\omega$,
(iii) the \textit{retarded region}: $\lambda_\omega \ll L$.
In zones (i) and (ii), retardation can be neglected (van-der-Waals zone),
while in zone (iii), it leads to a different power law (Casimir-Polder zone) for
the atom-surface interaction.

In the three regimes, different approximations for the reflection coefficients 
that appear in ~(\ref{eq:surface-Green-function})
can be made. In the \textit{sub-skin-depth zone} \cite{Henkel99}, we have $k\gg 
1/\delta_{\omega} \gg 1/\lambda_{\omega}$ and
% $k^2\gg\frac{\omega^2}{c^2}\epsilon(\omega)$
%
\begin{eqnarray}
r^{\rm TE}(\omega, k) &\approx& 
[ \epsilon(\omega) - 1 ] \frac{\omega^2}{4c^2k^2}~,
	\nonumber
\\
%%+[\epsilon^2(\omega)-1]\frac{\omega^4}{8c^4\kappa^4}+\cdots
r^{\rm TM}(\omega, k)&\approx& 
\frac{ \epsilon(\omega)-1 }{ \epsilon(\omega) + 1} 
\left[ 1+
\frac{\epsilon(\omega)}{\epsilon(\omega)+1}\frac{\omega^2}{c^2k^2}
\right]~.
	\label{eq:r_subskindepth}
%%+\frac{\epsilon^2(\omega)-\epsilon(\omega)}{\epsilon(\omega)+1}\frac
% {\omega^4}{4c^4\kappa^4}
\end{eqnarray}
At intermediate distances in the \textit{non-retarded zone},
the wave vector is $1 / \lambda_{\omega} \ll k \ll 1 / \delta_{\omega}$, hence
\begin{eqnarray}
r^{\rm TE}(\omega, k) &\approx& -1+\imath\frac{2}{\sqrt{\epsilon(\omega)}}
\frac{c k}{\omega},\nonumber\\
r^{\rm TM}(\omega, k) &\approx& 
1+\imath\frac{2}{\sqrt{\epsilon(\omega)}}\frac{\omega}{c k}~.
	\label{eq:r-non-retarded}
\end{eqnarray}
Finally, in the  \textit{retarded zone} we can consider $k\ll 1 / \lambda_{\omega} 
\ll 1/ \delta_{\omega}$, so that 
\begin{eqnarray}
r^{\rm TE}(\omega, k) &\approx& -1+\frac{2}{\sqrt{\epsilon(\omega)}}~,
\nonumber\\
r^{\rm TM}(\omega, k) &\approx& 1-\frac{2}{\sqrt{\epsilon(\omega)}}~.
\label{eq:epsilon_retarded}
\end{eqnarray}
Note that the first terms in Eqs.(\ref{eq:r-non-retarded}, \ref{eq:epsilon_retarded})
correspond to a perfectly reflecting medium (formally, $\varepsilon \to \infty$).
%
%A similar asymptotic analysis can be performed for the other model dielectric
%functions. It turns out that Eqs.(\ref{eq:r_subskindepth}--\ref{eq:epsilon_retarded})
%can still be used, provided the assumption $|\epsilon(\omega)| \gg 1$ holds.
%This is indeed the case for a typical atomic magnetic dipole moment and
%a conducting surface. 
%%For the plasma model, e.g., the role of 
%%the skin depth is played by the plasma wavelength $\lambda_{\rm p}$.

The asymptotics of the Green tensor that correlate to these
distance regimes 
are obtained by performing the $k$-integration in Eq.(\ref{eq:surface-Green-function})
with the above approximations for the reflection coefficients. The leading-order 
results
are collected in Table \ref{t:Green-function-asymptotes}. One 
notes that the $zz$-component is larger by a factor $2$ compared to
the $xx$- and $yy$-components. This difference
between the normal and parallel dipoles can be understood by the
method of images \cite{Jackson75}.

The magnetic response for a normally conducting metal in the sub-skin-depth 
regime is purely imaginary and scales linearly with the frequency $\omega$:
the 
reflected magnetic field is generated by induction. A significant response 
to low-frequency magnetic fields appears for superconductors because of the 
Meissner-Ochsenfeld effect \cite{Schrieffer99}.
In contrast, the electric response is 
strong for all conductors because surface charges screen the electric field 
efficiently.

The imaginary part of the trace of the Green tensor determines the 
local mode density (per frequency) for the electric or magnetic 
fields \cite{Joulain03}. These can be compared directly after 
multiplying by $\varepsilon_0$ (or $1/\mu_0$), respectively.
As is discussed in Refs.\cite{Joulain03, Henkel05a}, in the sub-skin-depth
regime near a metallic surface, the field fluctuations are mainly of 
magnetic nature. This can be traced back to the efficient screening 
by surface charges connected with electric fields.
Magnetic fields, however, cross the surface much more easily as surface 
currents are absent (except for superconductors). This reveals,
to the vacuum outside the metal, the thermally excited currents within the bulk.

%%######################################################
%\begin{table*}[tbh]
%\centering
%\begin{tabular}{l| c c c c c |c}
%\hline\hline
%& Sub -skin depth&\vline&Non-retarded &\vline& Retarded\\[1ex]
%&Drude 
%&%\vline %&plasma
%&\vline&\\
%\hline
%$\mathcal{H}_{xx}$ 
%&$\displaystyle \frac{\imath \mu_0}{32 \pi \delta^2_\omega L}$
%&\vline
%&
%&\vline
%& $\displaystyle - \frac{ \mu_0}{32 \pi L^3}$
%& $\displaystyle - \frac{ \mu_0}{32 \pi L^3} \left(1 - \frac{2 i \omega L}{c} - \frac{4 \omega^2 L^2}{c^2}\right)e^{2 i \omega L / c}$\\[1ex]
%$\mathcal{G}_{xx}$ 
%
%& 
%&
%& $\displaystyle \frac{1}{32\pi\epsilon_0 L^3}$ 
%&
%&
%& $\displaystyle \frac{1}{32\pi\epsilon_0 L^3}\left(1-\frac{2 \imath\omega L}{c}-\frac{4\omega^2 L^2}{c^2}\right)e^{2 \imath \omega L / c}$ \\
%\hline
%\end{tabular}
%\\
%\caption{Magnetic and electric Green tensors at a planar surface.
%The other elements have the asymptotes
%$\mathcal{H}_{yy} = \mathcal{H}_{xx}$,
%$\mathcal{H}_{zz} = 2\mathcal{H}_{xx}$, and similarly for
%$\mathcal{G}_{ii}$. The off-diagonal elements vanish.}
%\label{t:Green-function-asymptotes}
%\end{table*}
%%######################################################

\begin{table}[htdp]
\caption{Magnetic and electric Green tensors at a planar surface.
In this case the other elements have the asymptotes
$\mathcal{H}_{yy} = \mathcal{H}_{xx}$,
$\mathcal{H}_{zz} = 2\mathcal{H}_{xx}$, and similarly for
$\mathcal{G}_{ii}$. The off-diagonal elements vanish.
The expressions are for metals where $|\varepsilon(\omega) | \gg 1$.}
\centering
\begin{tabular}{c | c c c | c}
\hline\hline
 &\hspace*{.1cm}
 Sub-skin depth \hspace{-0.9cm} 
 &\vline&\hspace{-0.9cm} Non-retarded \hspace{.1cm}
 & Retarded
\\
\hline
 $\mathcal{G}_{xx}$
 &
 &\hspace*{-.5cm}
 $\displaystyle \frac{1}{32\pi\epsilon_0 L^3}
\Big( 1 - \frac{ 2 }{ \varepsilon(\omega ) } \Big)$ 
\hspace{-0.5cm} 
&\hspace{.2cm}& 
$\displaystyle \frac{1}{32\pi\epsilon_0 L^3}
\Big( 1 - \frac{ 2 }{ \varepsilon(\omega ) } \Big)
\left(1- \imath\frac{4\pi L}
{\lambda_{\omega}}-\frac{1}{2}\left[\frac{  4\pi L}{\lambda_{\omega}}\right]^
{2}\right)e^{4\pi \imath L / \lambda_{\omega}}$ 
\\[1ex]
$\mathcal{H}_{xx}$ 
 &\hspace*{-.2cm}$\displaystyle \frac{\imath \mu_0 }{32 \pi 
\delta^{2}_{\omega}  L}$\hspace{-0.5cm}
&\vline
&\hspace{-0.5cm} $
\displaystyle - \frac{ \mu_0}{32 \pi L^3}$ \hspace{.1cm} 
&$\displaystyle - \frac{ 
\mu_0}{32 \pi L^3} \left(1- \imath\frac{4\pi L}{\lambda_{\omega}}-\frac{1}{2}
\left[\frac{4\pi  L}{\lambda_{\omega}}\right]^{2}\right)e^{4\pi \imath L / 
\lambda_{\omega}}$\\
 \hline
\end{tabular}
\label{t:Green-function-asymptotes}
\end{table}%

\subsection{Asymptotic power laws}

To begin with, particle and field are both at zero temperature. The Matsubara 
series in Eq.\eqref{eq:Matsubara-series} can be replaced by an integral over 
imaginary frequencies:
\begin{equation}
\mathcal{F}= -\frac{\hbar}{2\pi}
\int\limits_{0}^{\infty}{\rm d}\xi
%\im\![\boldsymbol{\beta}^{T}(\omega) \cdot \boldsymbol{\mathcal{G}}(L, 
% \omega)]
\sum_{j}
\alpha_{jj}^{g}(\imath\xi)\mathcal{G}_{jj}(L, \imath\xi)~,
\label{zerotemp}
\end{equation}
where we have used the fact that the Green tensor is diagonal.  Alternatively one can get 
the previous result by taking the limit $T\to 0$ of 
Eq.\eqref{eq1:result-CP-free-energy} 
and performing a Wick rotation on the imaginary axis. One of the main 
advantages of this representation is that all functions in \eqref{zerotemp} are 
real. For electric dipole coupling, one has
$\alpha^{g}_{ii}(\imath \xi), \mathcal{G}_{ii}(L, \imath\xi) > 0$,
and we can conclude that the Eq.(\ref{zerotemp}) is a binding energy 
and corresponds to an attractive force 
(see the chapter by Capasso \emph{et al.} in this volume for detailed discussion on repulsive fluctuation-induced forces in liquids).

Along the imaginary axis, the Green tensor is dominated by an exponential
${\rm e}^{ - 2 \xi L / c }$, see Eq.(\ref{eq:surface-Green-function}).
This exponential 
suppresses large values of $\xi$ and the main contribution to the integral 
comes 
from the region $\xi < c/(2L)$. If this value is smaller than the characteristic 
frequency $\Omega_{e}$, say, of the atom or of the nanoparticle
(the lowest transition in eq. \eqref{eq:WS_polarizability}), 
the polarizability can be approximated by its static value.  
Assuming an isotropic polarizability we get the Casimir-Polder asymptote
($\lambda_{e }=c/(2\Omega_{e})$)
\begin{equation} 
\label{CasimirPolder}
L\gg \lambda_e: \quad
\mathcal{F}_{\rm CP}\approx
-\frac{3\hbar c \,\alpha^{g}_{\rm iso}(0) }{2^{5}\pi^{2}\epsilon_0 L^4}~,
\end{equation}
which is the well known expression for the atom-surface Casimir-Polder interaction 
\cite{Casimir48a}.
At short distance the polarizability limits the relevant frequency range to 
$\xi
\lesssim \Omega_{e}$. 
% and $\lambda_{\omega}\gtrsim \lambda_{e}$. 
Therefore 
for $L\ll\lambda_{e}$ we can replace Green tensor by its short distance 
approximation (see Table~\ref{t:Green-function-asymptotes})
where it becomes independent of $\xi$. We recover then the van 
der Waals asymptote
\begin{equation} 
\label{vanDerWaals}
L\ll\lambda_{e}: \quad
\mathcal{F}_{\rm vdW}\approx 
-\frac{\hbar}{2^{4}\pi^{2}
\epsilon_0 L^3}
\int\limits_{0}^{\infty}{\rm d}\xi
\, \alpha^{g}_{\rm iso}(\imath\xi)~.
\end{equation}
Similar expressions hold for the interaction due to a fluctuating magnetic 
dipole, the behaviour becoming more complicated when the distance
becomes comparable to a characteristic skin depth 
(see Eq.(\ref{eq:def-skin-depth}) and Ref.\cite{Haakh09b}). 
%
%in the retarded and non-
%retaded regime with the exception of the sub-skin depth region, where the behavior 
%is more complicated \cite{Haakh09b,Henkel05b}. Note, however, that the sign of 
%the magnetic atom-surface interaction is however reversed, corresponding to 
%repulsion (atomic paramagnetism).  We get again an attractive force for the 
%magnetic interaction between a surface and an nanosphere, because of 
%diamagnetic response of the nanoparticle.

%Let now rise the temperature. Since $\Omega_{e}\sim 10^{4}$K for $T=300 $K 
%we can reasonably approximate the electric thermal polarizability by its ground 
%state value. Thermal effects are then strongly correlated with the behavior of the 
%Green tensor. The temperature introduces another length scale
%\begin{equation}
%\lambda_{T}=\frac{\hbar c}{2\pi k_{B}T}
%\end{equation}
%which correspond to the average wave length of thermal photons. At room 
%temperature $\lambda_{T}$ is of the order of few micrometers which means that 
%thermal effects for the electric dipole have to be expected in the retarded region. 
If 
we write the Matsubara frequencies as $\xi_{n}= 2\pi n c/\lambda_{T}$
($n = 0, 1, 2, \ldots $),
the temperature may be low enough so that the limit $\lambda_{T}\gg L$ 
holds. Then all Matsubara frequencies 
are relevant, and if they are dense enough ($\lambda_{T}\gg \lambda_e$), 
the effect of temperature is negligible. The series in \eqref{eq:Matsubara-series} 
is then well approximated by the integral in \eqref{zerotemp}.
In the opposite (high-temperature) limit, one has $\lambda_{T}\ll L$ 
so that the exponential behavior of 
the Green tensor limits the series in \eqref{eq:Matsubara-series} to its first term recovering the Lifshitz asymptote
\begin{equation}
\label{eq:Matsubara-First}
\mathcal{F}_{\rm L}\approx -
  \frac{k_{B}T\, \alpha^{g}_{\rm iso}(0)}{16\pi\epsilon_0 L^3}~.
\end{equation}
We still have an attractive force. Note, however, that this attraction is mainly due 
to the classical part of the radiation, as the same result would be obtained with
a polarizable object immersed into the thermal field.

\section{Beyond equilibrium}
\label{s:non-equilibrium}

\subsection{Overview}

The theory presented so far has mainly considered atom, field, and surface to be
in a state of global thermal equilibrium, characterized by the same
temperature $T$. When one moves away from these conditions, the atom--surface
interaction assumes novel features like metastable or unstable states,
driven steady states with a nonzero energy flux etc. We review some of these
aspects here, since they have also appeared in recent 
experiments (Sec.\ref{sec:Exp}). On the theoretical side, there are a few
controversial issues that are currently under investigation \cite{Philbin09a,Pendry10,Leonhardt10}. 

We start with atoms prepared in non-thermal states: ground or excited
states that decay by emission or absorption of photons, 
and with atoms in motion where frictional forces appear. We then consider
field--surface configurations out of global equilibrium like a surface
surrounded by a vacuum chamber at different temperature. 

\subsection{Atoms in a given state and field in thermal equilibrium}
	\label{s:atoms-in-neq-state}
	
%\begin{equation}
%%\begin{split}
%\alpha_{ij}^a({\omega})=\sum_b
%\frac{d_i^{ab}d_j^{ba}}{\hbar}\frac{2\omega_{ba}}{\omega_{ba}^{2}
% - (\omega+\imath 0^{+})^{2}}
%%\end{split}
%\label{eq:WS_polarizability}
%\end{equation}
%{}[Eq.\eqref{eq:WS_polarizability} given below], 

The generalization of the Casimir-Polder potential 
to an atom in a definite state $|a\rangle$ 
can be found, for example, in 
Wylie \& Sipe \cite{Wylie85}, Eqs.(4.3, 4.4).
Now, the fluctuation--dissipation theorem for the dipole,
Eq.\eqref{eq:FTD_dipole}, does not apply, but perturbation theory is
still possible, with the result (summation over repeated indices
$i,j$)
\begin{multline}
	\label{eq:WS}
\mathcal{F}(L, T)=-k_{B}T \msum \alpha^{a}_{ij}(\imath \xi_n)\mathcal{G}_{ji}
(L,\imath \xi_n)
+\sum_b N(\omega_{ba}) d_i^{ab} d_j^{ba}
\re [\mathcal{G}_{ji}(L,\omega_{ba})]~,
\end{multline}
where $\overleftrightarrow{\alpha}^a$ is the state-dependent polarizability 
\cite{McLachlan63,Wylie85}. The dipole matrix elements are written
$d_i^{ab}=\langle a| {d}_{i} | b\rangle$.
The thermal occupation of photon modes (Bose-Einstein distribution) is
\begin{equation}
N(\omega)= \frac{ 1 }{ e^{\hbar\omega / k_{B} T } - 1}
= - 1 - N(- \omega )
	\label{eq:Bose-Einstein}~.
\end{equation}
Note the second term in Eq.(\ref{eq:WS})
that is absent in thermal equilibrium. It involves the
absorption and (stimulated) emission of photons on transitions
$a \to b$ to other quantum states, and the thermal occupation
number $N(\omega_{ba})$ evaluated at the Bohr frequency
$\hbar\omega_{ba} = E_b - E_a$. For this reason, it can be
called \emph{resonant} part. 
The first term that was also present
in equilibrium now features the state-dependent polarizability tensor
$\overleftrightarrow\alpha^a( {\rm i} \xi_n )$. This is the 
\emph{non-resonant} part of the interaction.

For the alkali atoms in their ground state $| a \rangle = | g \rangle$,
the Bohr frequencies $E_{bg}$ are all positive (visible and near-infrared
range) and much larger than typical laboratory temperatures (equivalent
to the THz range), hence the thermal occupation numbers $N(\omega_{ba})$
are negligibly small. By the same token, the ground-state polarizability 
is essentially the same as in thermal equilibrium 
$\overleftrightarrow\alpha^T( {\rm i} \xi_n )$ because the thermal
occupation of the excited states would come with an exponentially small
Boltzmann weight. The atom--surface interaction is then 
indistinguishable from its global equilibrium form and dominated by 
the non-resonant part. 

With suitable laser fields, one can perform the spectroscopy of
atom-surface interaction of excited states $| a \rangle = | e \rangle$.
It is also possible to prepare excited states by shining a resonant
laser pulse on the atom. 
In front of a surface, the second term in Eq.(\ref{eq:WS}) 
then plays a dominant role: the transition to the 
ground state where a real photon is emitted is accompanied by an
energy shift proportional to ${\rm Re} [\overleftrightarrow{\mathcal{G}}(L,
- \omega_{eg})]$. This resonant contribution 
can be understood in terms of the radiation reaction of a classical 
dipole oscillator \cite{Wylie85,Gorza06}: one would get the same result
by asking for the frequency shift of an oscillating electric dipole in front
of a surface -- a simple interpretation in terms of an image dipole is
possible at short distances (where the $k$-dependence of the reflection
coefficients~(\ref{eq:fresnel}) can be neglected). This term is essentially 
independent
of temperature if the transition energy $E_{eg}$ is above $k_B T$. 

A more familiar effect for the excited state is spontaneous decay,
an example for a non-stationary situation one may encounter out of
thermal equilibrium. 
We can interpret the resonant atom-surface interaction as the `reactive
counterpart' to this dissipative process. Indeed, the spontaneous decay
rate is modified relative to its value in free space by the presence of the 
surface. This can also be calculated in classical terms, leading to a
modification that involves the imaginary part of the Green tensor
$\overleftrightarrow G(L, \omega_{eg} )$. (The free-space 
contribution $\overleftrightarrow G_0$ has a finite imaginary part.)
In fact, both
the decay rate and the interaction potential can be calculated from a
complex self-energy of which Eq.(\ref{eq:WS}) is the real part in the 
lowest non-vanishing order of perturbation theory. 

What happens if the Bohr frequencies $\hbar\omega_{ba}$ become
comparable to $k_B T$? This applies, for example, to optically active
vibrational transitions and to atoms in highly excited states (Rydberg
atoms) where the energy levels are closely spaced. 
It is obvious from Eq.(\ref{eq:WS}) that the resonant term is subject to
cancellations among ``up'' ($E_{b} > E_{a}$) and
``down'' transitions ($E_{b'} < E_{a}$) with nearly degenerate 
Bohr frequencies: the occupation numbers $N( \omega_{ba} )$
and
$N( \omega_{b'a} )$ differ in sign, while 
${\rm Re}[ \overleftrightarrow{\mathcal{G}}( L, \omega )]$ is even
in $\omega$. To leading order in the high-temperature limit, the 
resonant term becomes
\begin{eqnarray}
	\mathcal{F}^{\rm res}(L,T) &\approx&
	\frac{ k_B T }{ \hbar }
	\Big[
	\sum_{b > a} \frac{ {\rm Re}\, \mathcal{G}^{ba}( L, \omega_{ba} ) 
	}{ \omega_{ba} }
	-
	\sum_{b' < a} \frac{ {\rm Re}\, \mathcal{G}^{b'a}( L, \omega_{ab'} ) 
	}{ \omega_{ab'} }
	\Big]~,
	\\
	\mathcal{G}^{ba}( L, \omega ) &=& d_i^{ab} d_j^{ba} 
	\mathcal{G}_{ij}( L, \omega )~,
	\label{eq:}
\end{eqnarray}
where the notation $b > a$ and $b < a$ means summing over states with
energies $E_b$ above or below $E_a$. 
This is proportional to the anharmonicity of the atomic level spectrum
around $E_a$. It vanishes exactly for a harmonic oscillator and reduces
significantly the coefficient linear in temperature in weakly anharmonic 
regions of the atomic spectrum \cite{Ellingsen10a}.

\subsection{Moving atoms}
	\label{s:moving-atoms}
	
An atom that moves in a radiation field can be subject to a frictional force,
as pointed out by Einstein in his seminal 1917 paper on the blackbody
spectrum \cite{Einstein17}. This force originates from the aberration and the Doppler shift
between the field the atom ``sees'' in its co-moving frame, and the
``laboratory frame''. (The latter frame is actually defined in terms of the
thermal distribution function of the radiation field that is not Lorentz-invariant.
Only the field's vacuum state in free space is Lorentz-invariant.) 
In addition, electric and magnetic fields mix under a Lorentz transformation
so that a moving electric dipole also carries a magnetic moment proportional
to ${\bf d} \times {\bf v}$ where ${\bf v}$ is the (center-of-mass) velocity
of the dipole (the R\"ontgen current discussed
in Refs.\cite{Wilkens93b, Wilkens94a, Wilkens94b, Scheel09}).

\subsubsection{Black body friction}

The free-space friction force ${\bf f}( {\bf v} , T )$ is given by
\cite{Mkrtchian03,ZuritaSanchez04}:
\begin{equation}
	{\bf f}( {\bf v}, T ) = - {\bf v} \frac{ \hbar^2 / k_B T
	}{ 12\pi^2 \varepsilon_0 c^5 }
	\int\limits_0^{\infty}\!{\rm d}\omega \frac{ \omega^5 \,
	{\rm Im}\,\alpha( \omega ) }{ \sinh^2(\hbar \omega / 2 k_B T) } ~,
	\label{eq:free-space-friction}
\end{equation}
where $\alpha( \omega )$ is the polarizability of the atom (in its
electronic ground state) and the approximation of slow motion (first order
in ${\bf v}/c$) has been made. For atomic transitions in the visible
range, this force is exponentially suppressed by the Boltzmann factor
$\sim {\rm e}^{ - \hbar \omega_{eg} / k_B T }$ that is winning
against the prefactor $1/T$ in Eq.(\ref{eq:free-space-friction}).
The physics behind
this effect is the same as in Doppler cooling in two counterpropagating
laser beams: the friction arises from the frequency shift in the frame
co-moving with the atom that breaks the efficiency of absorbing
photons with counter- and co-propagating momenta. Einstein derived the
Maxwell-Boltzmann distribution for the atomic velocities by balancing
this radiative friction with the momentum recoil in randomly distributed
directions as the absorbed photons are re-emitted, which leads to 
Brownian motion in velocity space \cite{Einstein17, Kleppner05}.
Conversely, assuming thermal equilibrium and the validity of the
Einstein relation between momentum diffusion and friction,
one can calculate the (linear) friction
tensor $\overleftrightarrow\Gamma$ 
in ${\bf f}( {\bf v} ) = - \overleftrightarrow\Gamma {\bf v}$
from the correlation function of the force 
operator \cite{ZuritaSanchez04, Schaich81}:
\begin{equation}
	\overleftrightarrow\Gamma = \frac{ 1 }{ k_B T }
	\int\limits_{-\infty}^{+\infty}\!{\rm d}\tau
	\langle {\bf F}( t +\tau ) {\bf F}( t ) \rangle~,
	\label{eq:friction-and-force-correlation}
\end{equation}
where ${\bf F}( t )$ is the force operator in the Heisenberg picture,
the operator product is symmetrized (as in Sec.\ref{sec:polarization-energies}),
and the average $\langle \cdots \rangle$ is taken at (global)
thermal equilibrium. One recognizes in Eq.(\ref{eq:friction-and-force-correlation})
the zero-frequency component of the force correlation spectrum.

The motion of atoms in the radiation field plays a key role for laser cooling
of ultracold gases. Although a discussion of laser-induced forces is beyond
the scope of this chapter, the basic principles can be illustrated by moving
away from global equilibrium and assigning temperatures $T_A$, $T_F$ 
to atom and field. An ultracold gas, immediately after switching off the 
lasers, would correspond to $T_A$ in the nanoKelvin range, while 
$T_F = 300\,{\rm K}$ is a good assumption for the fields in 
a non-cryogenic laboratory apparatus. Dedkov \& Kyasov calculated
the separate contributions from fluctuations of the atomic dipole
and the field, respectively. We follow here Ref.\cite{Dedkov09a}. 
Qualitatively speaking, the fluctuating dipole 
experiences a force when it emits a photon; this force is nonzero and depends
on velocity, even
after averaging over all emission directions, because the emission is
isotropic only in the rest frame of the atom. 
The absorption of photons from the fluctuating field is accompanied by
photon recoil, and here isotropy is broken because the Doppler shift
brings certain directions closer to the resonance frequency. (The same
principle is behind the so-called Doppler cooling in two counterpropagating
beams.) The sum of the two contributions takes the form
(adapted from Eq.(29) of Ref.\cite{Dedkov09a})
\begin{eqnarray}
	{\bf f}( {\bf v} ) &=& - 
	\frac{ \hbar }{ 4\pi\varepsilon_0 c \, \gamma_{\bf v} }
	\int\!\frac{ {\rm d}^3k }{ \pi^2 }
	(\hat {\bf k} \cdot \hat {\bf v}) 
	(\omega')^2 	{\rm Im}\,\alpha( \omega' ) 
	\left(
	N( \omega, T_F ) - N( \omega', T_A )
	\right)
	\label{eq:Dedkov-Kyasov09a}
	~,
\\
	\omega' &=& \gamma_{\bf v}( \omega + {\bf k} \cdot {\bf v} )~,
%	\bar n( \omega, T ) &=& 
%	\frac{ 1 }{ {\rm e}^{ \hbar \omega / k_B T } - 1 }
\end{eqnarray}
where $\gamma_{\bf v} = (1 - {\bf v}^2/c^2)^{-1/2}$ is the relativistic
Lorentz factor, and $\hat{\bf v}$, $\hat{\bf k}$ are unit vectors 
along the atom's velocity and the photon momentum.
The photon frequency in the ``blackbody frame'' (field temperature $T_F$)
is $\omega = c |{\bf k}|$, and
$N( \omega, T )$ is the Bose-Einstein distribution for a mode
of energy quantum $\hbar\omega$ at temperature $T$. The term with
$N( \omega, T_F )$ gives the force due to absorption of thermal
photons, while $N( \omega', T_A )$ gives the force due to 
dipole fluctuations. The absorbed power has to be calculated in the 
atom's rest frame: the energy $\hbar\omega'$ times the photon number 
provides the electric field energy density in this frame, and
the absorption spectrum ${\rm Im}\left[\omega' \alpha( \omega' )\right]$ 
must be evaluated at the Doppler-shifted frequency $\omega'$.
This shifted spectrum
also appears in the fluctuation-dissipation theorem~(\ref{eq:FTD_dipole})
now applied locally in the atom's rest frame, and determines the dipole
fluctuations. 

At equilibrium and in the non-relativistic limit, the difference between the 
Bose-Einstein distributions can be expanded to give
\begin{equation}
	N( \omega, T_F ) - N( \omega', T_A ) \approx
	- ({\bf k}\cdot{\bf v}) \partial_\omega N( \omega, T )
	=
	({\bf k}\cdot{\bf v}) \frac{ \hbar / k_B T }{
	4 \sinh^2(\hbar \omega / 2 k_B T ) }~,
	\label{eq:derivative-BE-distribution}
\end{equation}
and performing the angular integration, one recovers 
Eq.(\ref{eq:free-space-friction}). As another example, let us
consider a ground-state atom ($T_A = 0$) moving in a ``hot''
field $T_F > 0$ with a small velocity. We can then put $N(
\omega', T_A ) = 0$ since for free space photons, $\omega' > 0$
(the positive-frequency part of the light cone in the $(\omega,{\bf k})$
space is a Lorentz-invariant set). Expanding in the Doppler shift,
performing the angular integration and making a partial integration,
one arrives at
\begin{equation}
	{\bf f}( {\bf v} ) \approx {\bf v}
	\frac{ \hbar }{ 3\pi^2\varepsilon_0 c^5  }
	\int\limits_{0}^{\infty}\!{\rm d}\omega \,
	\omega^2 {\rm Im}\,\alpha^{g}( \omega ) 
	\partial_\omega[
	\omega^3 
	N( \omega, T_F )
	] ~.
	\label{eq:non-eq-free-space-friction}
\end{equation}
Note that the function $\omega^3 N( \omega, T_F )$ has a positive
(negative) 
slope for $\omega \lesssim k_B T_F / \hbar$ 
($\omega \gtrsim k_B T_F / \hbar$ ), respectively.
The velocity-dependent
force thus accelerates the particle, ${\bf v} \cdot {\bf f} > 0$,
if its absorption spectrum has a stronger weight at sub-thermal frequencies.
This may happen for vibrational transitions in molecules and illustrates the
unusual features that can happen in non-equilibrium situations. Drawing again
the analogy to laser cooling, the radiative acceleration corresponds to the
``anti-cooling'' set-up where the laser beams have a frequency 
$\omega > \omega_{eg}$ (``blue detuning''). Indeed, we have just found that
the peak of the thermal spectrum occurs on the blue side of the atomic 
absorption lines.

\subsubsection{Radiative friction above a surface}

Near a surface, the fluctuations of the radiation field are distinct from
free space, and are encoded in the surface-dependent Green function
$\overleftrightarrow{\mathcal{G}}(L, \omega )$, see 
Eqs.(\ref{eq:FTD_field}, \ref{splitting-G}). In addition, one has to take into account 
that the available
photon momenta differ, since also evanescent waves appear whose
$k$-vectors have components larger than $\omega/c$. 
All these properties can be expressed in terms of the electromagnetic
Green tensor, assuming the field in thermal equilibrium. Let us consider
for simplicity an atom with an isotropic polarizability tensor 
$\alpha_{ij} = \alpha\,\delta_{ij}$, moving at a non-relativistic
velocity ${\bf v}$. 
From Eq.(118) in 
Ref.\cite{Volokitin07}, one then gets a friction force
\begin{equation}
	{\bf f}( {\bf v} ) = - \frac{ \hbar }{ 2 \pi^2 \varepsilon_0 }
	\int\limits_0^{\infty}\!{\rm d}\omega
	\left( - \partial_\omega N( \omega, T )
	\right)
	{\rm Im}\,\alpha( \omega ) \,
(\hat{\bf v} \cdot \nabla )({\bf v} \cdot \nabla' ) 
	{\rm tr} \, {\rm Im}\, \overleftrightarrow{G}(
	{\bf r}, {\bf r}', \omega )~,
	\label{eq:Volokitin-surface-friction}
\end{equation}
where the spatial derivatives are taken with respect to the two position 
variables of the Green tensor, and ${\bf r} = {\bf r}' = {\bf r}_A(t)$ is
taken afterwards. This expression neglects
terms of higher order in $\alpha$ that appear in the self-consistent
polarizability~(\ref{eq:result-dressed-polarizability}); it 
can also be found from 
Eqs.(25,26) in Zurita Sanchez \emph{et al.} \cite{ZuritaSanchez04}.

Note that
friction is proportional to the local density of field states, encoded in
the imaginary part of the (electric) Green tensor 
$\overleftrightarrow{G}$. 
If the motion is parallel to a plane surface,
the result only depends on the distance $L$ and is independent of time.
The friction force is comparable in magnitude to the free-space 
result~(\ref{eq:Dedkov-Kyasov09a}) if the distance $L$ is comparable
or larger than the relevant wavelengths $c/\omega$: the derivatives
in Eq.(\ref{eq:Volokitin-surface-friction}) are then of the order
$(\hat{\bf v} \cdot \nabla )({\bf v} \cdot \nabla' ) \sim
| {\bf v} | (\omega/c)^2$. At sub-wavelength distances, the non-retarded 
approximation for the Green tensor can be applied 
(see Table~\ref{t:Green-function-asymptotes}), and the previous expression becomes of
the order of $| {\bf v} |/L^2$. The remaining integral is then similar to
the temperature-dependent part of the atom-surface interaction discussed
in Sec.\ref{s:cold-atom-hot-body} below.

An expression that differs from Eq.(\ref{eq:Volokitin-surface-friction})
has been found by Scheel and Buhmann \cite{Scheel09} who calculated
the radiation force on a moving atom to first order in the velocity,
and at zero temperature. Their analysis provides a splitting into
resonant and non-resonant terms, similar to Eq.(\ref{eq:WS}).
For the ground state, the friction force is purely non-resonant and contains
a contribution from the photonic mode density, similar to
Eq.(\ref{eq:Volokitin-surface-friction}), and one from the R\"ontgen 
interaction that appears by evaluating the electric field in the frame co-moving
with the atom. Another non-resonant friction force appears due to a 
velocity-dependent shift in the atomic resonance frequency, but it vanishes
for ground-state atoms and for the motion parallel to a planar surface.
The remaining friction force becomes in the non-retarded limit and for
a Drude metal
\begin{equation}
	{\bf f}( {\bf v} ) \approx - 
	\frac{ {\bf v} }{ 16 \pi \varepsilon_0 \, L^5 }
	\sum_{a > g} \frac{ |{\bf d}^{ag}|^2 \, \omega_s \Gamma_a }{ 
	(\omega_{ag} + \omega_s)^3 }~,
	\label{eq:Stefan-2-friction-force}
\end{equation}
where the sum $a > g$ is over excited states, the relevant 
dipole matrix elements are
$|{\bf d}^{ag}|^2 = \sum_i | {d}^{ag}_i |^2$,
$\Gamma_a$ is the radiative width of the excited state (which also depends
on $L$), and $\omega_s$ the surface plasmon resonance. 
The fluctuations of the
electromagnetic field are calculated here without taking into account
the ``back-action'' of the atom onto the medium (see Ref.\cite{Schaich81}
and the discussion below).
%, as discussed in Ref.\cite{Schaich81}.

We briefly mention that the behaviour of friction forces in the limit
of zero temperature (``quantum friction'')
has been the subject of discussion that is still
continuing (see also the chapter by Dalvit \emph{et al.}\ in this volume for further discussion on quantum friction). 
An early result of Teodorovich on the friction force between two plates,
linear in ${\bf v}$ with a nonzero coefficient as $T \to 0$ \cite{Teodorovich78}, 
has been challenged by Harris and Schaich \cite{Schaich81}. They
point out that a charge
or current fluctuation on one metallic plate can only dissipate by exciting 
electron-hole pairs in the other plate, but the cross-section for this 
process vanishes like $T^2$. This argument does not hold, however,
for Ohmic damping arising from impurity scattering. In addition, 
Ref.\cite{Schaich81} points out that the fluctuations of the atomic dipole
should be calculated with a polarizability that takes into account the
presence of the surface. This self-consistent polarizability has been
discussed in Sec.\ref{sec:non-perturbative-level-shift} and reduces the
friction force, in particular at short (non-retarded) distances. Carrying
out the calculation for a metallic surface and in the non-retarded regime,
Harris \& Schaich find the scaling
\begin{equation}
	{\bf f}( {\bf v} ) \approx - {\bf v} 
	\frac{ \hbar \, \alpha_{\rm fs}^2 }{ L^{10} }
	\left( \frac{ \alpha(0) c }{ 4\pi\varepsilon_0 \omega_s } \right)^2 
	\label{eq:Harris-Schaich-non-retarded-friction}~,
\end{equation}
where $\alpha_{\rm fs}$ is the fine structure constant,
$\alpha(0)$ is the static polarizability of the atom, $e$ the
elementary charge, and $\omega_s = \omega_p / \sqrt{2}$ 
the surface plasmon frequency in the non-retarded limit. 
Note the different scaling with distance $L$ compared to 
Eq.(\ref{eq:Stefan-2-friction-force}).

\subsection{Nonequilibrium field}
\label{s:non-equilibrium-field}

A radiation field that is not in thermal equilibrium is a quite natural
concept since under many circumstances, an observer is seeing 
radiation where the Poynting vector is nonzero (broken isotropy) and 
where the frequency spectrum is not given by the (observer's)
temperature. The modelling of these fields can be done at various
levels of accuracy: ``radiative transfer'' is a well-known example from
astrophysics and from illumination engineering -- this theory can be 
understood as a kinetic theory for a ``photon gas''. It is, in its simplest
form, not a wave theory and therefore not applicable to the small
length scales (micrometer and below) where atom-surface interactions
are relevant. ``Fluctuation electrodynamics'' is a statistical description based 
on wave optics, developed by the school of S. M. Rytov \cite{Rytov3,Landau9}, 
and
similar to optical coherence theory developed by E. Wolf and co-workers
\cite{Mandel95}. The main idea is that the radiation field is generated
by sources whose spectrum is related to the local temperature and the
material parameters of the radiating bodies. The field is calculated by 
solving the macroscopic Maxwell equations, where it is assumed that the
matter response can be treated with linear response theory (medium
permittivity or dielectric function $\varepsilon( {\bf x}, \omega )$ and 
permeability $\mu( {\bf x}, \omega )$). This framework has been
used to describe the quantized electromagnetic field, as discussed by 
Kn\"oll and Welsch and their co-workers \cite{Scheel00b},
by the group of Barnett \cite{Barnett97}, see also the review paper
Ref.\cite{Henry96}. Another application is radiative heat transfer and
its enhancement between bodies that are closer than the thermal
(Wien) wavelength, as reviewed in Refs.\cite{Joulain05a, Volokitin07}.
The non-equilibrium heat flux between two bodies at different temperatures
is naturally calculated from the expectation value of the Poynting vector.

In this section we review the atom-surface interaction in the out-of-equlibrium configuration similar to the one studied in \cite{Antezza05}: the atom is close to a substrate hold at temperature $T_S$, the whole being enclosed in a  ``cell'' with walls (called "environment") at temperature $T_E$. In the following we will only consider the electric atom-surface interaction and we will use the zero temperature atomic electric polarizability. In fact, the electric dipole transitions are mainly in the visible range and their equivalent in temperature ($10^{3 - 4}$K) are not achieved in the experiment. Therefore the atom does not participate in the thermal 
exchange and can be considered in its ground state.

\subsubsection{Fluctuation electrodynamics and radiative forces}

A very simple non-equilibrium situation occurs when an atom is located
near a ``heated body'' whose temperature is larger than its ``surroundings''.
\begin{SCfigure}
%\begin{center}
\centering
\includegraphics*[width=75mm]{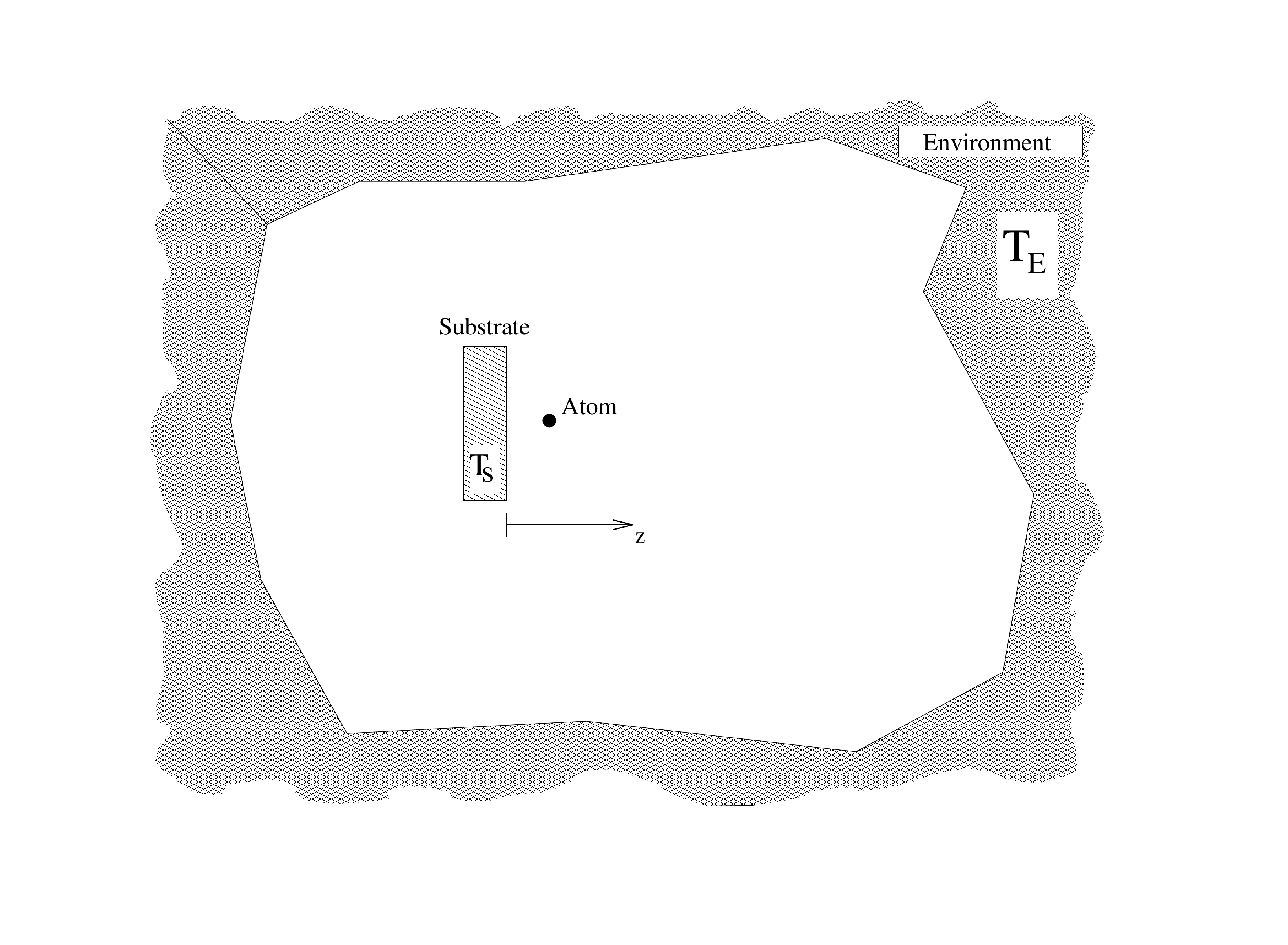}
\caption{Sketch of an atom-surface system with the field being out of
thermal equilibrium. $T_S$ is the temperature of the substrate and $T_E$ is the temperature of the walls of the cell surrounding the atom-substrate system. If $T_S > T_E$, there is a nonzero radiative heat flux
from the surface into the surrounding environment.
\label{fig:configuration_neq}}
%\end{center}
\end{SCfigure}
As mentioned above, it is quite obvious that the Poynting vector of the
radiation field does not vanish: there is radiative heat flux from the body
into the surrounding space. This flux is accompanied by a radiative force
on the atom that depends on the atomic absorption spectrum, but also
on the angular distribution of radiated and re-scattered photons. The 
atom-photon interaction, in this case, does not derive from the gradient of
a potential. The basic concept is that of the radiation force ${\bf F}$; 
it is given by \cite{Jackson75}
\begin{equation}
	{\bf F}( {\bf r} ) = \langle d_i \nabla E_i( {\bf r} ) \rangle
 	+ \langle \mu_i \nabla B_i( {\bf r} ) \rangle
	+ \mbox{ higher multipoles }
	\label{eq:radiation-force-in-dipole-approximation}~,
\end{equation}
where we have written out only the contributions from the electric
and magnetic dipole moments and the atom is assumed at rest at
position ${\bf r}$. (The generalization to a moving atom leads to the
velocity-dependent forces discussed in Sec.\ref{s:moving-atoms}.)
As a general rule, the electric dipole is the 
dominant contribution for atoms whose absorption lines are in the visible
range. Eq.(\ref{eq:radiation-force-in-dipole-approximation}) can be derived
by averaging the Coulomb-Lorentz force over the charge and current
distribution in the atom, assuming that the atomic size is small compared
to the scale of variation (wavelength) of the electromagnetic field.
The average $\langle \ldots \rangle$ is taken with respect to the
quantum state of atom and field, and operator products are taken in
symmetrized form.

The radiation force~(\ref{eq:radiation-force-in-dipole-approximation}) can
be evaluated with the scheme outlined in Sec.\ref{sec:polarization-energies}
where the operators ${\bf d}$ and ${\bf E}( {\bf r} )$ are split into
``freely fluctuating'' and ``induced'' parts. Carrying this through for 
the contribution of field fluctuations, leads to an expression of the form
%\begin{equation}
%	\langle d_i^{{\rm free}} \nabla E_i^{{\rm in}} \rangle =
%%	
%	\label{eq:}
%\end{equation}
\begin{equation}
	\langle d_i^{{\rm in}} \nabla E_i^{{\rm free}}( {\bf r} ) \rangle =
	\int\!\frac{ {\rm d}\omega }{ 2\pi }
	 \,\frac{ {\rm d}\omega' }{ 2\pi } \, \alpha_{ij}( \omega )
	\langle E_j^{{\rm free}}( {\bf r}, \omega ) 
	\nabla E_i^{{\rm free}}( {\bf r}, \omega' )
	\rangle
	\label{eq:force-from-field-fluctuations}~,
\end{equation}
where the spatial gradient of a field autocorrelation function appears.
In a non-equilibrium situation, the fluctuation-dissipation theorem of 
Eq.(\ref{eq:FTD_field}) cannot be applied, and this field correlation must 
be calculated in a different way. In a similar way, one gets
\begin{equation}
	\langle d_i^{{\rm free}} \nabla E_i^{{\rm in}}( {\bf r} ) \rangle =
	\int\!\frac{ {\rm d}\omega }{ 2\pi }
	 \,\frac{ {\rm d}\omega' }{ 2\pi } \, 
	\nabla_1 G_{ij}( {\bf r}, {\bf r}, \omega' )
	\langle d_i^{{\rm free}}(  \omega ) 
	d_j^{{\rm free}}( \omega' )
	\rangle
	\label{eq:force-from-dipole-fluctuations}~,
\end{equation}
where the gradient $\nabla_1 G_{ij}$ is evaluated with respect to the
first position coordinate of the Green function. This term requires some
regularization because of the divergent Green function at coincident 
positions. The correlation function of the atomic dipole 
can be calculated in its stationary state which could be a thermal
equilibrium state or not, as discussed in Sec.\ref{s:atoms-in-neq-state}.
In the case of an ultracold atomic gas, it is clear that the atom can be at
an effectively much lower temperature compared to the macroscopic
bodies nearby. This is consistent with the perturbation theory behind the 
operator splitting into fluctuating and induced parts. In global equilibrium,
when both fluctuation spectra are given by the fluctuation-dissipation
theorem~(\ref{eq:FTD_field}, \ref{eq:FTD_dipole}), it can be seen easily that
the force reduces to the gradient of the equilibrium interaction 
potential~(\ref{eq1:result-CP-free-energy}).

Within Rytov's fluctuation electrodynamics, the fluctuating field is given in 
terms of its sources and the macroscopic Green function. Generalizing
Eq.(\ref{eq:field-G-and-dipole}), one gets
\begin{equation}
	E_i( {\bf r}, \omega ) = 
	\int\!{\rm d}^3r'\,
	G_{ij}( {\bf r}, {\bf r}', \omega ) 
	P_j( {\bf r}', \omega )
	+ \mbox{magnetization sources}
	\label{eq:Rytov-representation-with-polarization-noise}~,
\end{equation}
where we have not written down the contribution from the magnetization
field that can be found in Ref.\cite{Buhmann07a}.
The polarization density $P_j( {\bf r}', \omega )$ describes the
excitations of the material (dipole moment per unit volume). 
If the material is locally stationary, 
the polarization operator averages to zero 
and its correlations
$\langle P_i( {\bf r}, \omega ) 
	 P_j( {\bf r}', \omega' ) \rangle$
determine the field spectrum
\begin{equation}
	\langle E_i( {\bf r}, \omega ) E_j( {\bf r}', \omega' ) \rangle
	= 
	\int\!{\rm d}^3x\,{\rm d}^3x'\,
	G_{ik}( {\bf r}, {\bf x}, \omega )
	G_{jl}( {\bf r}', {\bf x}', \omega' )
	\langle P_k( {\bf x}, \omega ) 
	 P_l( {\bf x}', \omega' ) \rangle
	\label{eq:fields-correlations-from-local-source}~,
\end{equation}
Making the key assumption of local thermal equilibrium at the
temperature $T( {\bf r}) $, the source correlations are 
given by the \emph{local version} of the fluctuation-dissipation 
theorem~\cite{Rytov3}:
\begin{equation}
	\langle P_i( {\bf r}, \omega ) 
	 P_j( {\bf r}', \omega' ) \rangle = 
	 2\pi\hbar \,
	 \delta( \omega + \omega' )
	 \delta_{ij}
	 \delta( {\bf r} - {\bf r}' )
	 \coth\Big(\frac{ \hbar \omega }{ 2 k_B T({\bf r} ) }\Big)
	{\rm Im}\,[ \varepsilon_0\varepsilon( {\bf r}, \omega ) ]
	\label{eq:FDT-local-sources}~,
\end{equation}
where $\varepsilon( {\bf r}, \omega )$ is the (dimensionless)
dielectric function of the
source medium, giving the polarization response to a local
electric field. The assumption of a local and isotropic (scalar)
dielectric function explains the
occurrence of the terms
$\delta_{ij} \delta( {\bf r} - {\bf r}' )$; this would not apply
to ballistic semiconductors, for example, and to media with spatial 
dispersion, in general. The local temperature 
distribution $T( {\bf r} )$ should in that case be smooth on the length
scale associated with spatial dispersion (Fermi wavelength, screening
length, mean free path). Similar expressions for random sources
are known as ``quasi-homogeneous sources'' and are studied in
the theory of partially coherent fields \cite[\S5.2]{Mandel95}.

If we define a polarization spectrum by
\begin{equation}
	S_P( {\bf r}, \omega ) = \hbar 
	 \coth\Big(\frac{ \hbar \omega }{ 2 k_B T({\bf r} ) }\Big)
	{\rm Im}\,[ \varepsilon_0\varepsilon( {\bf r}, \omega ) ]
	\label{eq:def-polarization-spectrum}~,
\end{equation}
the field correlation function~(\ref{eq:fields-correlations-from-local-source}) 
becomes
\begin{equation}
	\langle E_i( {\bf r}, \omega ) E_j( {\bf r}', \omega' ) \rangle
	= 
	2 \pi \delta( \omega + \omega' ) 
	\int\!{\rm d}^3x\,
	G_{ik}^*( {\bf r}, {\bf x}, \omega' )
	G_{jk}( {\bf r}', {\bf x}, \omega' )
	S_P( {\bf x}, \omega' )
	\label{eq:result-field-correlations}~,
\end{equation}
where
Eq.(\ref{symmetry}) has been applied to the Green function.
This expression was named ``fluctuation-dissipation theorem of the second kind''
by Eckhardt \cite{Eckhardt82} who analyzed carefully its limits of
applicability to non-equilibrium situations.
Note that even if the sources are spatially decorrelated (different points
${\bf r}$ and ${\bf r}'$ in Eq.(\ref{eq:FDT-local-sources}), the 
propagation of the field creates spatial coherence, similar to the lab class
diffraction experiment with a coherence slit. The spatial coherence
properties of the field determine the order of magnitude of the gradient
that is relevant for the radiation force in Eq.(\ref{eq:force-from-field-fluctuations}).

Let us focus in the following on the correction to the atom-surface
force due to the thermal radiation created by a ``hot body''. 
We assume that the atom is in its ground state and evaluate the
dipole fluctuation spectrum in Eq.(\ref{eq:force-from-dipole-fluctuations})
at an atomic temperature $T_A = 0$. 
To identify the non-equilibrium part of the force, it is useful to split
the field correlation spectrum in Eq.(\ref{eq:force-from-field-fluctuations})
into its zero-temperature part and a thermal contribution, using
$\coth( \hbar \omega / 2 k_B T ) = {\rm sign}(\omega)
[1 + 2 N( |\omega|, T ) ]$ with the
Bose-Einstein distribution $N( \omega, T )$. The Rytov currents are
constructed in such a way that at zero temperature, 
the fluctuation-dissipation theorem~(\ref{eq:FTD_field}) for the field
is satisfied. This can be achieved by allowing formally for a nonzero
imaginary part ${\rm Im}\,\varepsilon( {\bf r}, \omega ) > 0$ everywhere
in space \cite{Scheel00b,Eckhardt84}, or by combining the radiation of 
sources located
inside a given body and located at infinity \cite{Henry96, Savasta2002}.
The two terms arising from dipole and field fluctuations
then combine into a single
expression where one recognizes the gradient of the Casimir-Polder
potential Eq.(\ref{eq1:result-CP-free-energy}). This is discussed in
detail in Refs.\cite{Henkel02, Antezza08}. 
The remaining part of the atom-surface force that depends on the
body temperature is discussed now.

\subsubsection{Radiation force near a hot body}
	\label{s:cold-atom-hot-body}
	
Let us assume that the body has a homogeneous temperature
$T( {\bf x} ) = T_S$ and a spatially constant dielectric function.
% in Eq.(\ref{eq:def-polarization-spectrum}). 
Using Eq.(\ref{eq:result-field-correlations}) and subtracting the $T = 0$
limit, the spectrum of the 
nonequilibrium radiation (subscript `neq') can then be expressed 
by the quantity
\begin{eqnarray}
	\langle E_i( {\bf r}, \omega ) E_j( {\bf r}', \omega' ) 
	\rangle_{\rm neq}
	&=&
	2\pi \delta( \omega + \omega' ) 
	N( |\omega| , T_S ) 
	\hbar  \, 
	S_{ij}( {\bf r}, {\bf r}', \omega )~, \qquad
\\
	S_{ij}( {\bf r}, {\bf r}', \omega )
	&=&
	{\rm Im}\,[ \varepsilon_0 \varepsilon( |\omega| ) ]
	\int\limits_{S}\!{\rm d}^3x\,
	G_{ik}^*( {\bf r}, {\bf x}, \omega )
	G_{jk}( {\bf r}', {\bf x}, \omega )
	\label{eq:def-Wij-thermal-correlation}~,
\end{eqnarray}
where the space integral is over the volume of the body. The tensor
$S_{ij}( {\bf r}, {\bf r}', \omega )$ captures the material composition
of the body and its geometry relative to the observation points.

\begin{figure}[hbt]
%\begin{center}
\includegraphics*[width=55mm]{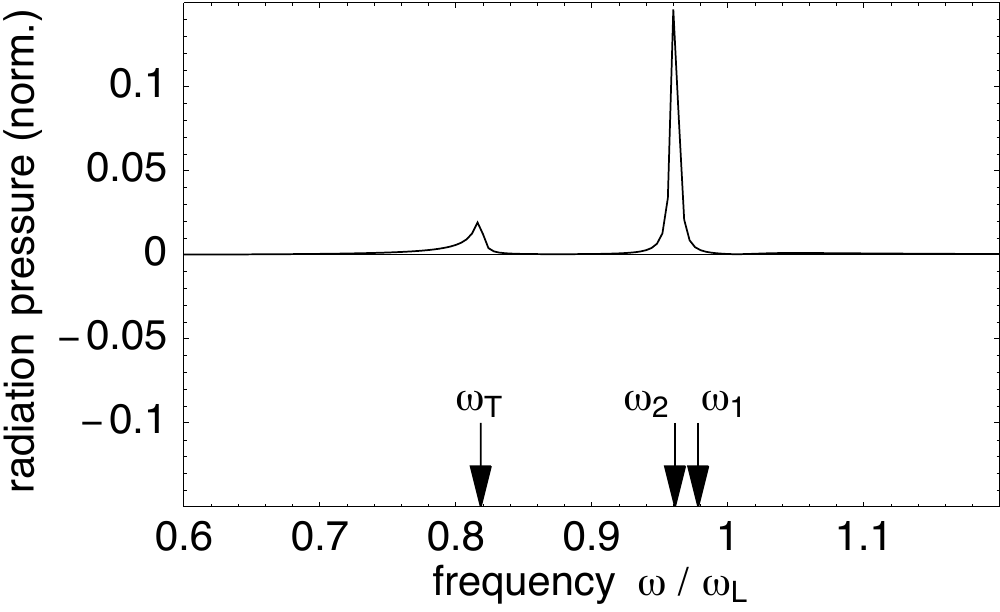}
\includegraphics*[width=60mm]{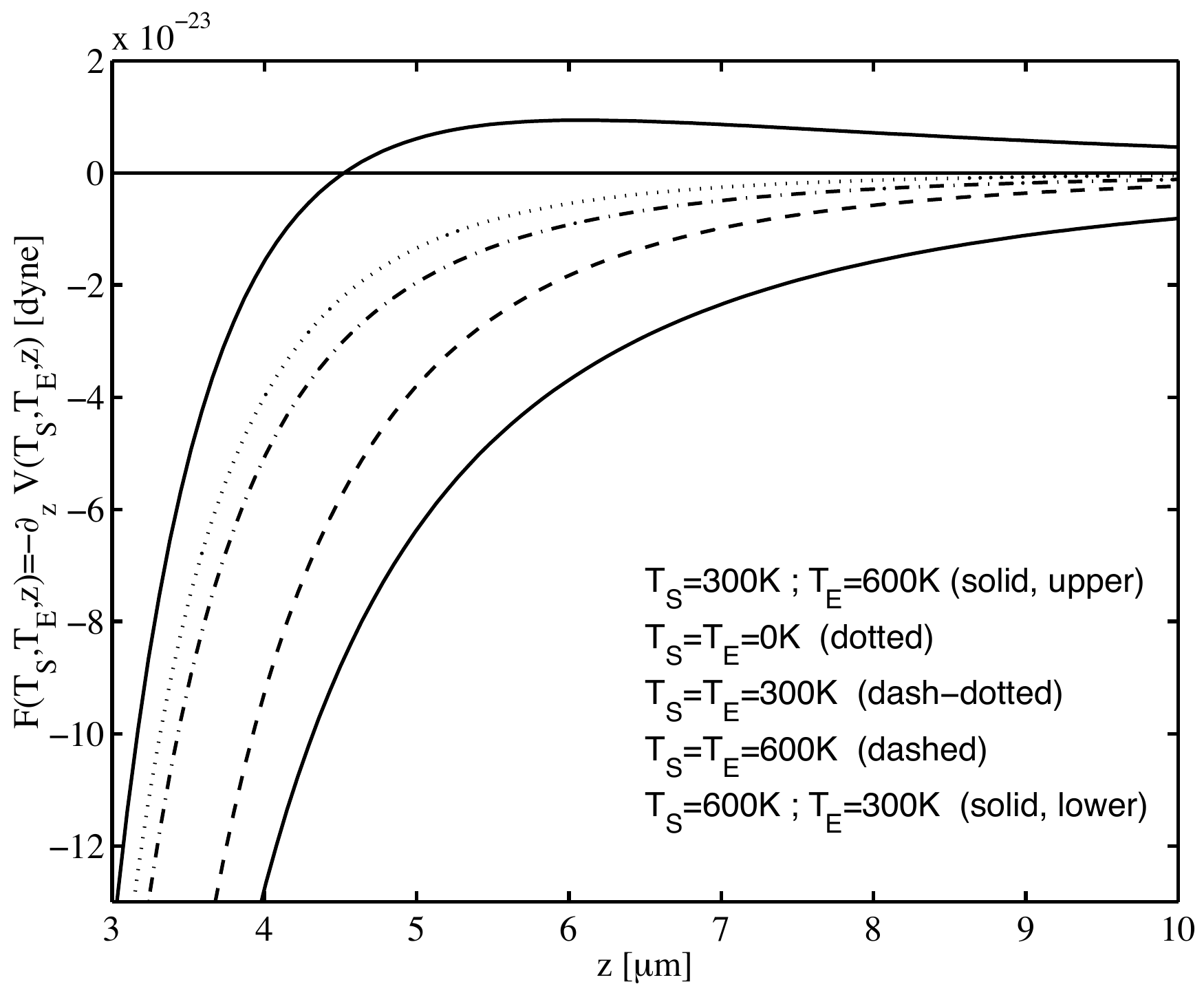}
%\end{center}
\caption[]{(left) Spectrum of thermal radiation pressure 
exerted on a small spherical particle above a planar substrate
(positive = repulsive, does not depend on distance).
The arrows mark the substrate and particle resonances at
${\rm Re}\,\varepsilon^{-1}( \omega_{\rm T} ) = 0$ 
(bulk phonon polariton) 
and 
${\rm Re}\,\varepsilon( \omega_{1,2} ) = -1,-2$ 
(surface and particle phonon polariton).
The force spectrum is given by $\hbar N( \omega, T_S )$ times
the second term in Eq.(\ref{eq:split-the-force}), and
normalized to $(16/3) \hbar k_{L} (k_{L} a)^{3} N( \omega, T_S )$ 
where $a$ is the particle radius and $k_{L} = \omega_{L} / c$
the wavenumber of the longitudinal bulk polariton
(${\rm Re}\,\varepsilon( \omega_{\rm L} ) = 0$).
\\
(right)
Theoretical calculation of the atom-surface force, in and out of thermal 
equilibrium, taken from Ref.\cite{Antezza05}, Fig.2. 
The atom is Rubidium 87 in its electronic ground state, the surface is made of
sapphire (SiO$_2$). Note the strong variation of the
non-equilibrium force, both in magnitude and in sign. A negative sign
corresponds to an attractive interaction.
	\label{fig:Force_NEQ}}
	\label{fig:radiation-pressure}
\end{figure}

Referring to the force due to field fluctuations in 
Eq.(\ref{eq:force-from-field-fluctuations}), let us assume for simplicity
that the atomic polarizability is isotropic, 
$\alpha_{ij}( \omega ) = \delta_{ij} \alpha( \omega )$. 
We combine the integrand over positive and negative frequencies 
to isolate dispersive and absorptive contributions ($\omega > 0$)
\begin{eqnarray}
	&&
	\alpha^*( \omega ) \nabla_2 S_{ii}( {\bf r}, {\bf r}, \omega )
	+ \mbox{$( \omega \mapsto - \omega )$}
	\nonumber
\\
	&&
	=
	{\rm Re}[ \alpha( \omega ) ]
	\nabla_{\bf r} [S_{ii}( {\bf r}, {\bf r}, \omega )] 
	+ 
	2\, {\rm Im}[ \alpha( \omega ) ]
	\,{\rm Im}[ 
	\nabla_{2} S_{ii}( {\bf r}, {\bf r}, \omega ) 
	] 
	\label{eq:split-the-force}~,
\end{eqnarray}
where $\nabla_2$ is the gradient with respect to the second coordinate
of $S_{ij}$, while $\nabla_{\bf r}$ differentiates both coordinates. This
form highlights that the non-equilibrium force separates in two \cite{Chaumet00,Arias-Gonzalez03} contributions
that are familiar in laser cooling \cite{Dalibard85a,Cohen-Tannoudji98}: a \emph{dipole force} equal to
the gradient of the electric energy density (proportional to
$S_{ii}( {\bf r}, {\bf r}, \omega ) \ge 0$). 
This is proportional to the real part of $\alpha$ 
and can be interpreted as the polarization energy of the atom in the
thermal radiation field. The second term in Eq.(\ref{eq:split-the-force})
gives rise to \emph{radiation pressure}, it is generally\footnote{Note that when the dressed polarizability is 
used instead of the bare one, the polarizability has an imaginary part even in the absence of absorption (see discussion at the end of section \ref{sec:non-perturbative-level-shift}). This is equivalent to include the effects of the ``radiative reaction'' in the dynamic of the dipole \cite{Draine88,Chaumet00,Milonni10a} as required by the conservation of energy and the optical theorem.} proportional to the atomic
absorption spectrum and the phase gradient of the field. 
The phase gradient
can be identified with the local momentum of the emitted photons. 
By inspection of Eq.(\ref{eq:def-Wij-thermal-correlation}) for a planar
surface, one indeeds
confirms that the force pushes the atom away from the thermal source. 
An illustration is given in Fig.\ref{fig:radiation-pressure} for a 
nanoparticle above a surface, both made from semiconductor. The
dielectric function $\varepsilon( \omega )$ is of Lorentz-Drude form 
and uses parameters for SiC
(see Ref.\cite{Henkel02}). The arrows mark the resonance frequencies
of transverse bulk phonon polaritons $\omega_{\rm T}$ and of 
the phonon polariton 
modes of surface ($\omega_1$) and particle ($\omega_2$).

The radiation pressure force is quite difficult to observe with atomic
transitions in the visible range because the peaks of the absorption
spectrum are multiplied by the exponentially small Bose-Einstein factor
$N( \omega_{ag}, T_S )$, even if the body temperature reaches the
melting point. 
Alternative settings suggest polar
molecules or Rydberg atoms \cite{Ellingsen10a} with lower transition
energies.
In addition, some experiments are only sensitive to force
gradients (see Sec.\ref{sec:Exp}), and it can be shown that the 
radiation pressure above a planar surface (homogeneous temperature,
infinite lateral size) does not change with distance. 

For the evaluation of the dipole force, the same argument related to the
field temperature can be applied
so that the atomic polarizability in Eq.(\ref{eq:split-the-force})
is well approximated by its static value,
${\rm Re}\,\alpha( \omega ) \approx \alpha( 0 )$.
We thus find
\begin{eqnarray}
	{\bf F}_{\rm neq}^{\rm dip}( {\bf r}, T_S,T_E=0)&=& - \nabla 
	U_{\rm neq}^{\rm dip}( {\bf r}, T_S,T_E=0)~,
	\label{eq:result-dipole-force1}
\\
	U_{\rm neq}^{\rm dip}( {\bf r}, T_S,T_E=0) &=& -
	\alpha( 0 )
	\int\limits_0^\infty\!\frac{ {\rm d}\omega }{ 2\pi }
	\hbar N( \omega, T_S)
	S_{ii}( {\bf r}, {\bf r}, \omega )	
	\label{eq:result-dipole-force2}~.
\end{eqnarray}

\subsubsection{General non equilibrium configuration and asymptoric behaviours}

In the general case both $T_S$ and $T_E$ can be different from zero \cite{Antezza05}. The total force will be the sum 
%of the force at zero-temperature  (derived from Eq.\eqref{zerotemp}), and of a pure thermal component. The latter is the sum 
of $\mathbf{F}^{\rm{dip}}_{\rm{neq}}(\mathbf{r},T_S,T_E=0)$ given in the previous expression and of 
\begin{equation}
\mathbf{F}^{\rm{dip}}_{\rm{neq}}(\mathbf{r},T_S=0,T_E)=\mathbf{F}^{\rm{dip}}_{\rm{eq}}(\mathbf{r},T_E)-\mathbf{F}^{\rm{dip}}_{\rm{neq}}(\mathbf{r},T_E,T_S=0),
\end{equation}
that is the difference between thermal force at equilibrium at the temperature $T_E$ and the force in Eqs.\eqref{eq:result-dipole-force1} and \eqref{eq:result-dipole-force2} with $T_{S}$ and $T_{E}$ swapped.
An illustration of the resulting force is given in Fig.\ref{fig:Force_NEQ}(right)
for a planar surface at temperature $T_S>T_{E}$. 
%The equilibrium force is included
%in this calculation, and $T_E$ means the temperature
%of the surroundings.
A large-distance asymptote of the non-equilibrium interaction can be derived in the 
form \cite{Antezza05,Antezza06a,Pitaevskii06}
\begin{equation}
	L \gg \frac{ \lambda_T }{ \sqrt{ \varepsilon(0) -1 } }: \qquad
	U^{\rm dip}_{\rm neq}( L, T_S, T_E )  \approx - \frac{ \pi }{ 12 }
	\alpha( 0 ) \frac{ \varepsilon(0) + 1}{ \sqrt{ \varepsilon(0) - 1 } }
	\frac{ k_B^2 (T_S^2 - T_E^2 ) }{ \hbar c \, L^2 }
	\label{eq:non-eq-asymptote}~,
\end{equation}
where $\varepsilon( 0 ) < \infty$ is the static dielectric constant.
The previous expression is valid for dielectric substrates, see 
Ref.\cite{Antezza05} for Drude metals.
For $T_S = T_E$, this formula vanishes, and one ends with the ``global 
equilibrium'' result of Eq.(\ref{eq:WS}). 

Expression \eqref{eq:non-eq-asymptote} shows that the configuration out of thermal equilibrium presents new features with respect to the equilibrium
force. Indeed the force scales as the difference of the square of the temperatures and can be attractive or repulsive. For $T_E>T_S$ the force changes sign, going from attractive at small distance to repulsive at large distance (i.e. featuring a unstable equilibrium position in between), and it decays slower than the equilibrium configuration ($\propto L^{-4}$), leading therefore to a stronger force. This new feature was crucial for the first measurement of the thermal component of the surface-atom surface (see next section). Moreover, when a gas of atoms is placed in front of the surface, the non-equilibrium interaction can lead to interesting non-additive effects \cite{Antezza06,Antezza08}.

\section{Measurements of the atom-surface force with cold atoms}
	\label{sec:Exp}

\subsection{Overview}

We do not discuss here the regime of distances comparable to the atomic
scale where atomic beams are diffracted and reveal the crystallography
of the atomic structure of the surface. We consider also that the atoms are kept away even from being 
physisorbed in the van der Waals well (a few nanometers above the surface).
One is then limited to distances above approximately one micron (otherwise
the attractive forces are difficult to balance by other means), but can take
advantage of the techniques of laser cooling and micromanipulation
and use even chemically very reactive atoms like the alkalis. 

The first attempts to measure atom-surface interactions in this context
go back to the sixties, using atomic beam set-ups. 
In the last 20 
years, technological improvements have achieved the 
sensitivity required to detect with good accuracy and precision tiny forces.
As examples, we mention
the exquisite control over atomic beams provided by laser cooling \cite{Sandoghdar92, 
Sukenik93}, spin echo techniques that reveal the quantum reflection of
metastable noble gas beams
\cite{Shimizu01, Druzhinina03} (see also the chapter by DeKieviet \emph{et al.} for detailed discussions on this topic),
or the trapping of an ultracold laser-cooled gas in atom chip devices
\cite{Vuletic04, Obrecht07}. 
In this section we will 
briefly review some of the experiments which exploited cold atoms in order to 
investigate the Casimir-Polder force. 

By using different techniques, it has been possible to measure the atom-surface 
interaction (atomic level shift, potential, force, or force gradient, depending on the 
cases) both 
at small ($0.1\,\mu{\rm m} < L < \lambda_{\rm{opt}} \approx
0.5\mu{\rm m} $) and large ($L > 1\mu{\rm m}$) 
distances.
Due to the rapid decrease of the interaction as the atom-surface separation 
becomes larger, the small-distance 
(van der Waals-London, Eq.\eqref
{vanDerWaals}) regime at $L < \lambda_{\rm{opt}}$ is somewhat
easier to detect. Recall that in this limit, 
only the vacuum fluctuation of electromagnetic field are relevant, and retardation 
can be ignored. More recent experiments explored the weaker
interaction in the Casimir-Polder regime (\ref{CasimirPolder}),
$\lambda_{\rm{opt}} < L < \lambda_T$,  
where retardation effects are relevant,
but thermal fluctuations still negligible. Also the 
Lifshitz-Keesom regime at $L > \lambda_T$ has been explored,
where thermal fluctuations are dominant 
[see Eq.\eqref{eq:Matsubara-First}].  
The theory of Dzyaloshinskii, Lifshitz, and Pitaevski  (DLP, \cite[Ch.VIII]{Landau9})
encompasses the three regimes with crossovers that are illustrated in 
Fig.\ref{fig:potential-overview} for a Rubidium atom and a room temperature
sapphire surface.

\subsection{From van der Waals to Casimir-Polder: equilibrium}

Typically, experiments have been performed at room temperature, and at thermal 
equilibrium, and used several techniques to measure the interaction, usually of 
mechanical nature. 

The van der Waals-London regime has been first explored by its effect on the 
deflection of an atomic beam passing close to a substrate \cite
{Raskin69,Shih74,Shih74a,Shih75,Anderson86}. Such kind of experiments were 
almost qualitative, and hardly in agreement with the theory. Subsequently, more 
accurate measurements of the atom-surface interaction in this regime have been 
done by using dielectric surfaces ``coated'' with an
evanescent laser wave that repels the atoms (atom mirrors, \cite{Landragin96}),
by atom diffraction from transmission gratings \cite{Grisenti99,Bruhl02}, 
by quantum 
reflection \cite{Shimizu01,Druzhinina03}, and by spectroscopic studies \cite
{Sandoghdar92,Fichet07}.

\begin{figure}[htb]
\includegraphics*[width=55mm]{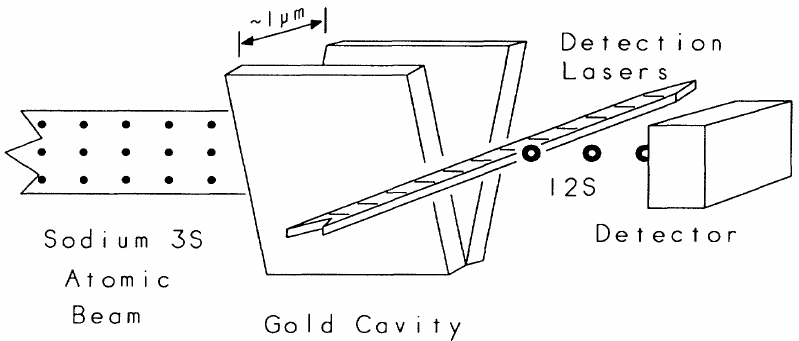}
\includegraphics*[width=60mm]{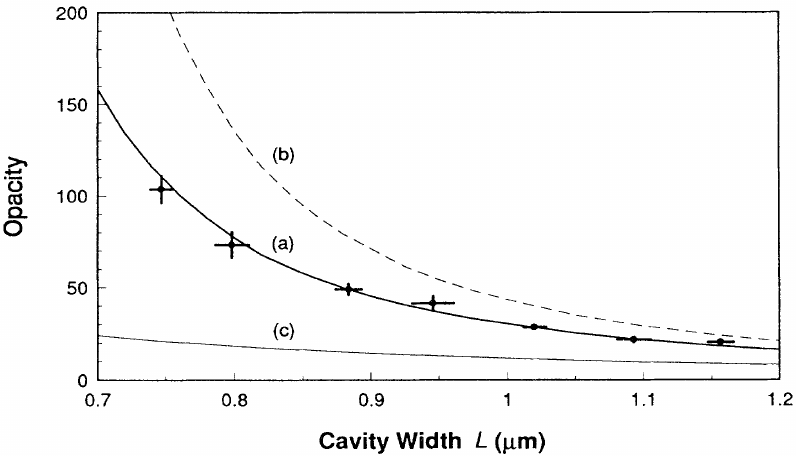}
\caption[]{(left)
Scheme of the experiment of Sukenik \& al, taken from Ref.\cite{Sukenik93}.
An atomic beam enters a micron-sized gold cavity, and the flux of atoms 
emerging the cavity is detected and related to the atom-surface potential inside 
the cavity. \label{fig:Hinds1}
(right)
Measurement of the atom-surface interaction in the Casimir-Polder regime, 
in the experiment of \cite{Sukenik93}, taken from the same paper. The opacity is 
proportional to the number of atoms which do not exit from the cavity, and is 
related to the atom-surface potential. The solid lines are the theoretical prediction 
based on: (a) full DLP potential, (b) van der Waals-London (short-distance) potential, 
and (c) no atom-surface potential. 
}
\label{fig:Hinds2}
\end{figure}

The Casimir-Polder regime, where vacuum fluctuations of the 
electromagnetic field and the finite speed of light are 
relevant, was first studied experimentally in  \cite{Sukenik93}\footnotemark[2]. Here the force has been 
measured through an atomic beam deflection technique, which consists in letting
an atomic beam (Na atoms in their ground state) pass across in a cavity made 
by two 
walls (gold plates), as one can see in fig.\ref{fig:Hinds1}. The atoms of the beam 
are drawn by the Casimir-Polder force to the walls, whose intensity 
depends on the atomic position within the cavity.
Part of the atoms are 
deflected during their path and stick to the cavity walls without reaching the end 
of the cavity. Only few atoms will pass the whole cavity, and their flux is measured 
and related to the atom-surface interaction in the cavity. 
Such a measurement is shown 
in fig.~\ref{fig:Hinds2}, where the theoretical curves are based on atomic trajectories
in the atom-surface potential that is assumed to be either of van der Waals-London
or of Casimir-Polder form. The data are clearly consistent with the CP interaction,
hence retardation already plays a role for typical distances in the range of 
$500\,{\rm nm}$. 

\begin{SCfigure}
%\begin{center}
\centering
\includegraphics*[width=60mm]{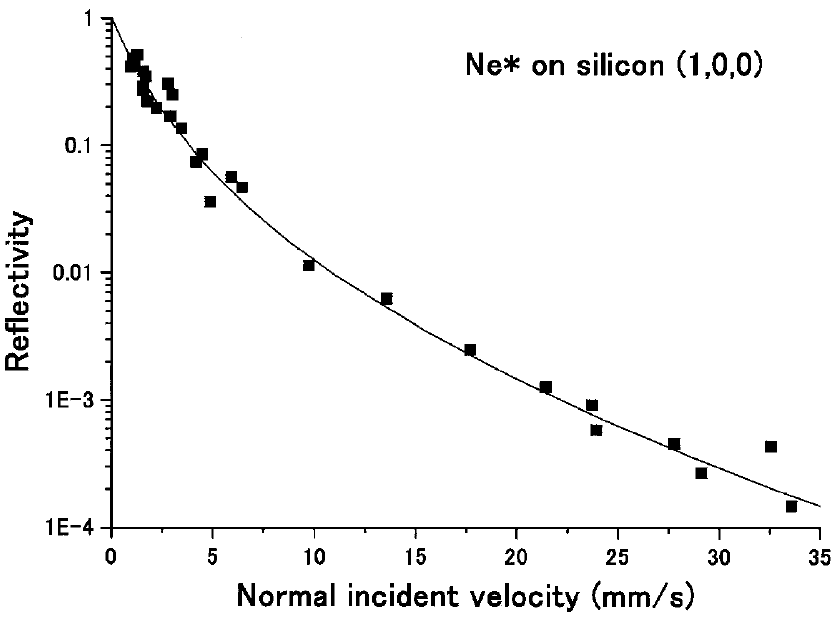}
%\end{center}
\caption{Reflectivity as a function of the normal incident velocity of Ne* atoms on 
a Si(1,0,0) surface, taken from Fig.3 of \cite{Shimizu01}. The experimental points 
(squares) are plotted together with a theoretical line calculated using the 
approximate expression $V_{\rm CP}=-C_4/[(d+a)d^3]$, where $C_4=6.8\;10^
{-56}$ Jm$^4$ and $a$ is a fitting parameter.}
\label{fig:Shimizu}
\end{SCfigure}

Subsequent 
measurements of the Casimir-Polder force have been done, among others,  using
the phenomenon of quantum reflection of ultra-cold atomic beams from a solid
surface \cite{Shimizu01, Druzhinina03}. In the experiment,
a collimated atomic beam of metastable Neon atoms impinges on
a surface (made of Silicon or some glass) at a glancing angle (very small 
velocity normal to the surface). In this regime, the de Broglie wave of the
incident atoms must adapt its wavelength to the distance-dependent potential,
and fails to do so because the potential changes too rapidly on the scale of the
atomic wavelength. This failure forces the wave to be reflected, a quantum effect
that would not occur in an otherwise attractive potential. In the limit of zero
normal velocity (infinite wavelength), the reflection probability must reach 100\%.
The variation with velocity depends on the shape of the atom-surface potential
and reveals retardation effects \cite{Cote98, Friedrich02}.
In fig.~\ref
{fig:Shimizu} is shown the measurement of quantum reflection performed by 
Shimizu \cite{Shimizu01}\footnotemark[3].
In this case, the accuracy was not high enough to distinguish reliably between
theoretical predictions. More recent data are shown in Fig.5 of the chapter by DeKieviet {\it et al.} in this volume.
\footnotetext[2]{Reprinted figures with permission from 
C.~I. Sukenik, M.~G. Boshier, D. Cho, V. Sandoghdar, and E.~A. Hinds, \href{http://prl.aps.org/abstract/PRL/v70/i5/p560_1}{Phys. Rev. Lett. {\bf 70},  560
(1993)}. Copyright (1993) by the American Physical Society.}

\footnotetext[3]{Reprinted figure with permission from 
F. Shimizu, \href{http://prl.aps.org/abstract/PRL/v86/i6/p987_1}{Phys. Rev. Lett. {\bf 86},  987  (2001)}. Copyright (2001) by the American Physical Society.}

\begin{SCfigure}
%\begin{center}
\centering
\includegraphics*[width=60mm]{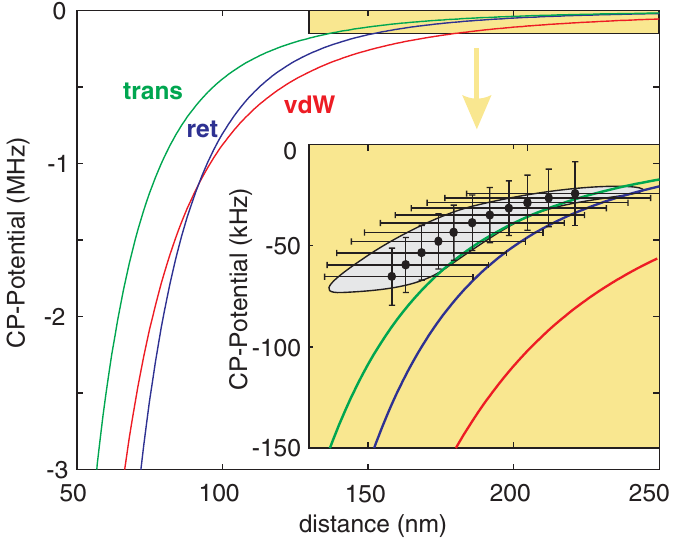}
\caption{Measured and theoretical prediction for the Casimir-
Polder interaction, taken from fig.3 of \cite{Slama10}. In the large figure theoretical 
calculation: asymptotic van der Waals-London (vdW), asymptotic  Casimir-Polder 
(ret), and full theoretical curve (trans). In the inset,  measured data points are 
included: statistical and systematic errors are indicated by the error bars,
and the gray shaded area, respectively. \label{fig:Slama}}
%\end{center}
\end{SCfigure}

The crossover between the van der Waals and Casimir-Polder regimes 
has been recently measured by the group of J. Fort\'agh \cite{Bender10}, 
using the reflection of a cloud of
ultracold atoms at an evanescent wave atomic mirror. This experiment improves
previous data obtained by the A. Aspect group \cite{Landragin96} into the
crossover region. The data are shown in Fig.\ref{fig:Slama} where 
``vdW'' and ``ret'' label the asymptotes van der Waals and Casimir-Polder potentials,
respectively\footnote[4]{Reprinted figure with permission from 
H. Bender, P.~W. Courteille, C. Marzok, C. Zimmermann, and S. Slama, \href{http://prl.aps.org/abstract/PRL/v104/i8/e083201}{Phys.
  Rev. Lett. {\bf 104},  083201  (2010)}. Copyright (2010) by the American Physical Society.}. The full calculation (DLP theory)
is labelled ``trans'' and shows some deviation
in the crossover region. The data (shown with error bars) are clearly favoring the 
theory including retardation.

%\clearpage

\subsection{The Cornell experiments}

\subsubsection{Lifshitz regime}

\begin{figure}[htb]
\includegraphics*[width=55mm]{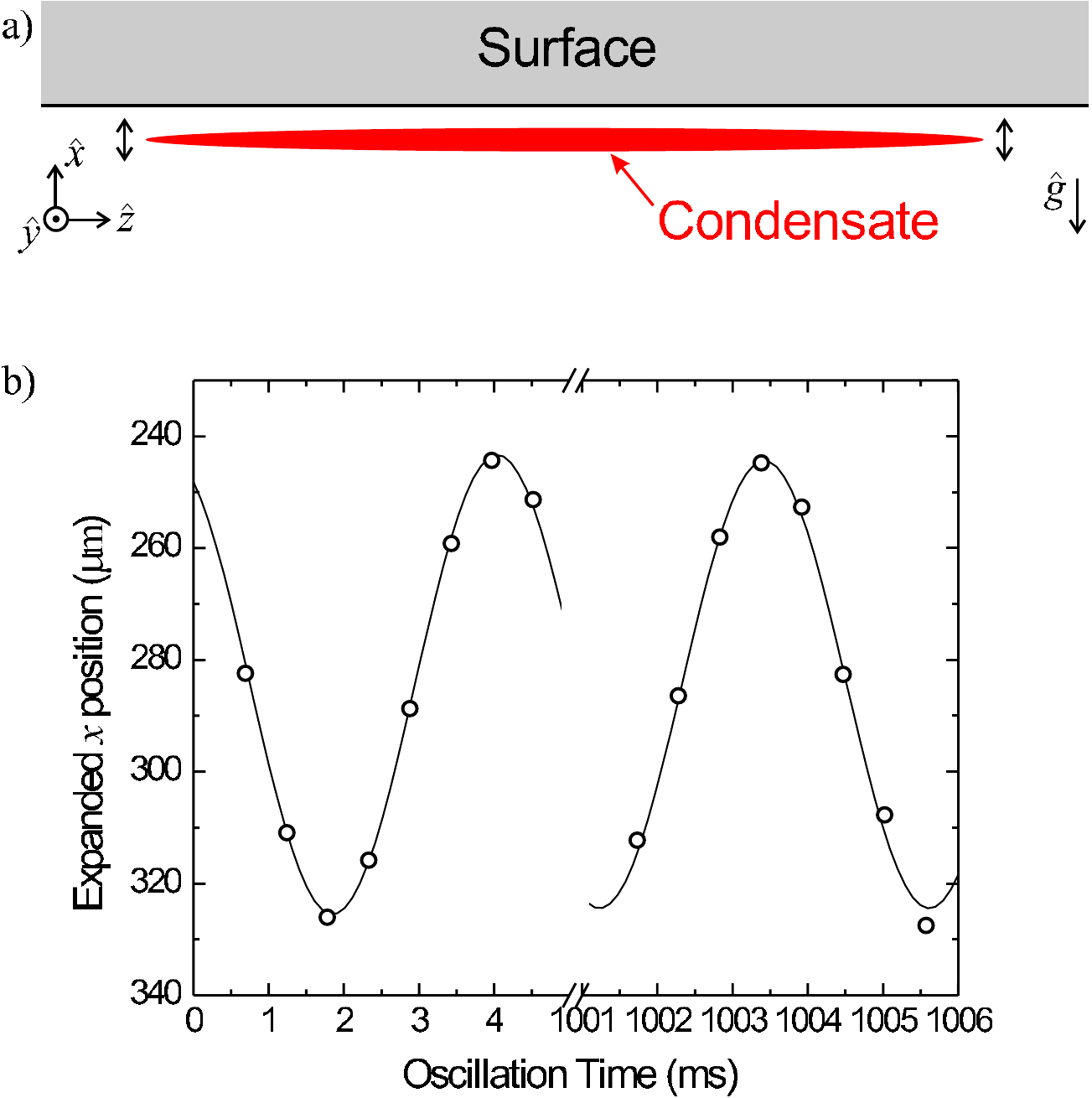}
\includegraphics*[width=60mm]{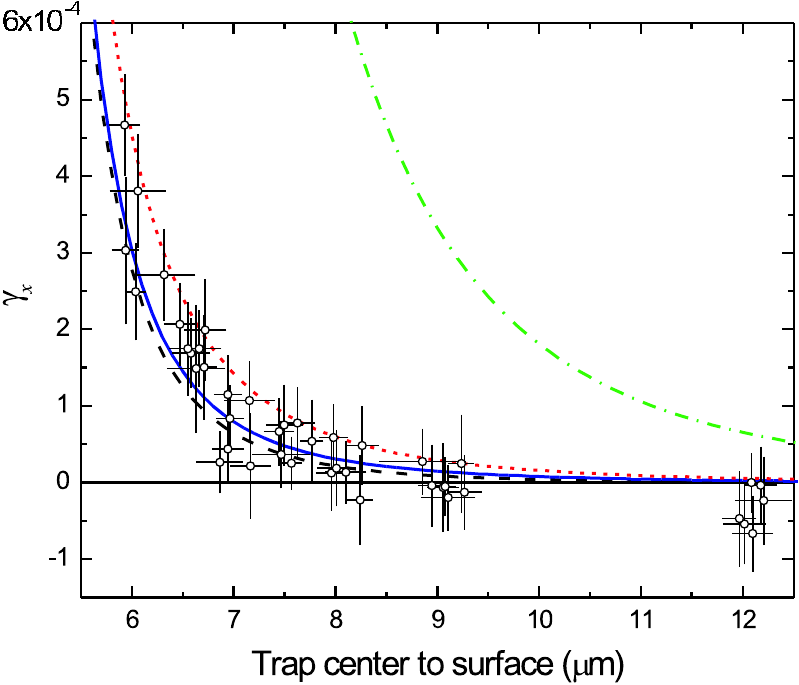}
\caption[]{(left)
Scheme illustrating the experimental configuration in the E. Cornell group,
taken from Fig.1 of \cite{Harber05}.
(a) Typical arrangement of the condensate close to the surface. The cloud is formed
by a few hundred thousand Rubidium~87 atoms, its axial length is 
$\sim 100\,\mu{\rm m}$. The surface is made of fused silica.
The coordinate axes and the direction of gravity are indicated.  
(b) Typical data showing the center-of-mass (dipole) 
oscillation ($x$-direction normal to the surface).
This is obtained after holding the BEC near the surface and then
shifting it rapidly away from the surface; the
``expanded position'' is proportional to the velocity component $v_x$.
\label{fig:Cornell1}
\\
(right)
Measured and theoretical frequency of the BEC center-of-mass motion,
relative to the nominal trap frequency
$\omega_{\rm trap}$ and normalized as $\gamma = (\omega_{\rm cm} -
\omega_{\rm trap})/\omega_{\rm trap}$.
Each data
point represents a single measurement, with both statistical and systematic errors. 
The mean oscillation amplitude is $\approx 2.06 \mu{\rm m}$, and the typical
size of the BEC (Thomas-Fermi radius) in the oscillation direction is 
$\approx 2.40 \mu $m.
Theory lines, calculated using theory from \cite{Stringari04}, consider the full atom-
surface potential: $T=0$ K (dashed, black line), $T=300$ K (solid, blue line) and 
$T=600$ K (dotted, red line).  No adjustable
parameters have been used. The result of the van der Waals-London potential has 
been added (dash-dotted, green line).}
\label{fig:Cornell2}
\end{figure}

The atom-surface interaction in the Lifshitz regime has been explored 
in Cornell's group \cite{Harber05, Obrecht07}.
Here a quantum degenerate gas in the 
Bose-Einstein condensed phase has been used as local sensor to 
measure the atom-surface interaction, similar to the work in V. Vuleti\'c's group
where smaller distances were involved \cite{Vuletic04}.
The Cornell experiments use a Bose-Einstein condensate (BEC) of a few
$10^5$ $^{87}$Rb atoms that are
harmonically trapped at a frequency $\omega_{\rm trap}$.
The trap is moved towards
the surface of a sapphire substrate, as illustrated in Fig.\ref{fig:Cornell1}.

Center-of-mass (dipole) oscillations of the trapped gas are then excited in the 
direction normal to the surface.  In absence of atom-surface interaction, the 
frequency of the center-of-mass oscillation would correspond to the frequency of 
the trap: $\omega_{\rm cm}=\omega_{\rm trap}$. 
Close to the substrate, the atom-surface potential changes the effective trap frequency, shifting  the center-of-mass frequency by a quantity $\gamma=(\omega_{\rm cm}-\omega_{\rm trap})/\omega_{\rm trap}$. The value of $\gamma$ is related to the atom-surface force \cite{Antezza04}  and for small oscillation amplitudes we have:
 \begin{equation}
\omega_{\rm cm}^2=\omega_{\rm trap}^2+\frac{1}{m}\int_{-{R_z}}^{{R_z}}dz\;n_0^z(z)\; \partial^{2}_{z} \mathcal{F}(z)
\label{eq:relative-frequency-shift}
\end{equation}
where $m$ is the atomic mass, $\mathcal{F}(z)$ is the atom-surface free energy  in Eq. \eqref{eq1:result-CP-free-energy} ($z$ is the direction normal to the surface) , and $n_0^z(z)$ is the $xy$-integrated unperturbed atom density profile \cite{Antezza04} that takes into account the finite size of the gas cloud. In the Thomas-Fermi approximation 
\begin{equation}
n_0^z(z)=\frac{15}{16R_z}[1-\left(\frac{z}{R_z}\right)^2]^2,
\end{equation}
where $R_z$ (typically of few microns) is the cloud radius along $z$, which depends on the chemical potential. In the comparison with the experiment, non-linear effects due to large oscillation amplitudes \cite{Antezza04} may become 
relevant \cite{Harber05}.
%%Close to the substrate, the  second derivative of the atom-surface potential changes the effective 
%%trap frequency, shifting the center-of-mass frequency by a quantity 
%\begin{equation}
%	\gamma = \frac{ \omega_{\rm cm} - \omega_{\rm trap} }{
%	\omega_{\rm trap} } \approx
%	\frac{ \overline{ \partial_L^2 U( L ) } }{ 2 m \omega_{\rm trap}^2 }
%	\label{eq:relative-frequency-shift}~.
%\end{equation}
%%This frequency shift must take into account the finite size of the cloud
%%(indicated by the overbar), and non-linear effects due to large-amplitude
%%oscillations, as studied in Ref.\cite{Antezza04}.
The experiment of Ref.\cite{Harber05} was performed at room temperature and
succeeded in measuring the atom-surface interaction for the first time up 
to distances $L \approx 7\mu{\rm m}$
(see Fig.\ref{fig:Cornell2}). 
Although the relative frequency shift in Eq.(\ref{eq:relative-frequency-shift})
is only $\sim 10^{-4}$, the damping of this dipole
oscillator is so weak that its phase can be measured even after hundreds
of periods, see Fig.\ref{fig:Cornell2}(left).
The same technique has been recently proposed to test the interaction between an 
atom and a non-planar surface \cite{Dalvit08,Messina09}.

% \clearpage
 
\subsubsection{Temperature dependence and non-equilibrium force}

\begin{SCfigure}%[htb]
\includegraphics*[width=55mm]{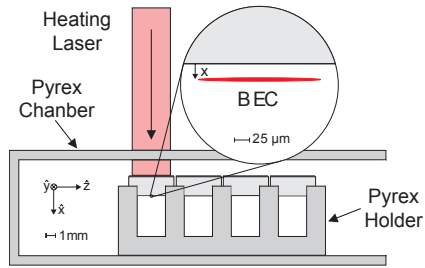}
\caption[]{
Scheme of the experiment of Ref.\cite{Obrecht07} (from which the 
figure is taken), where atom-surface interactions out of thermal equilibrium
have been measured.}
\label{fig:Cornell3}
\end{SCfigure}

\begin{SCfigure}%[htb]
\includegraphics*[width=60mm]{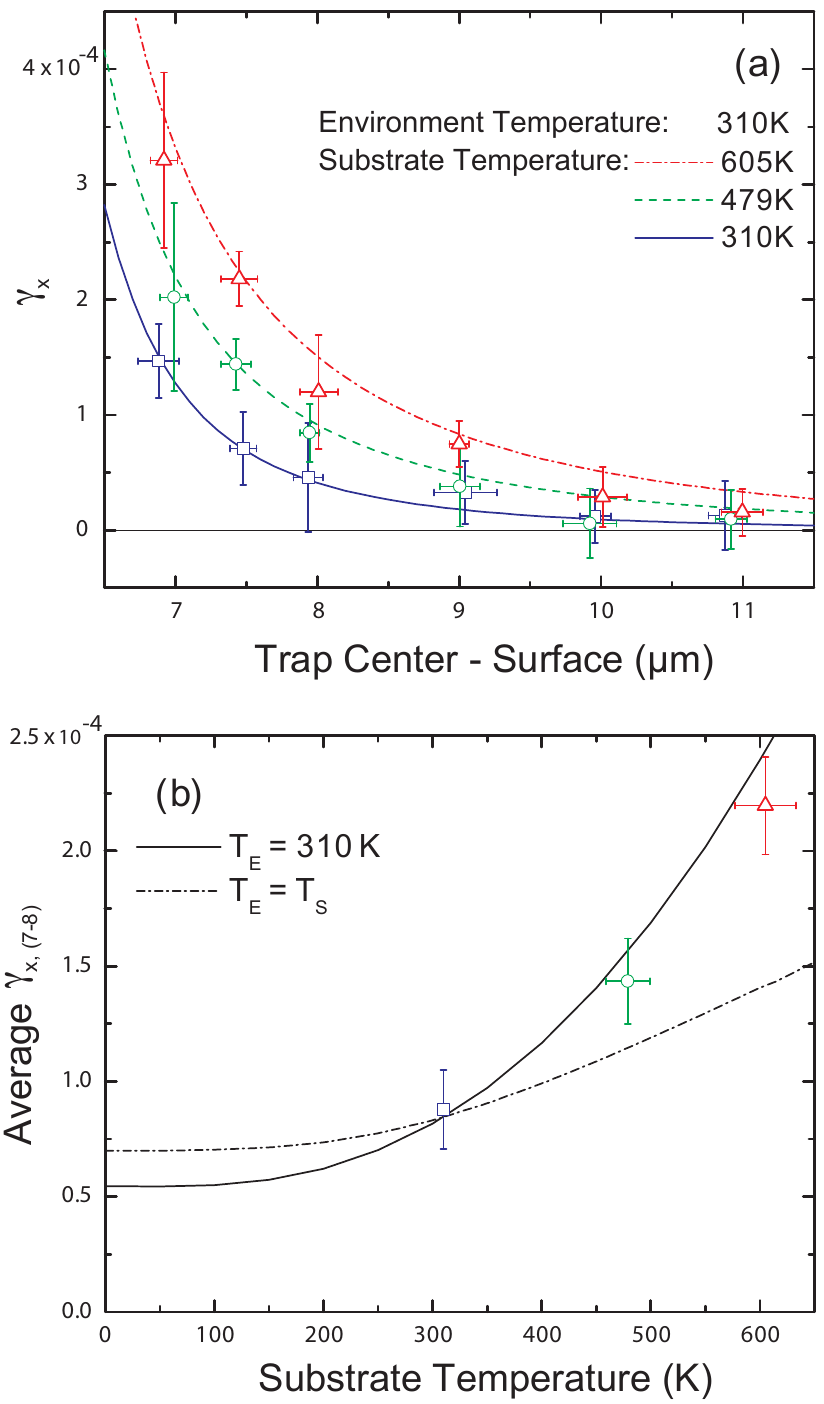}
\caption[]{
Measured and theoretical frequency shift $\gamma$ of the center-of-
mass frequency $\omega_{\rm cm}$ for a trapped atomic BEC ($^{87}$Rb 
atoms) close to surface (fused silica), in a system in and out of thermal equilibrium. 
The figure is taken from fig.4 of \cite{Obrecht07}.
(a) The figure shows three sets of
data and accompanying theoretical curves with no adjustable
parameters for various substrate temperatures. Data are shown for different 
substrate temperatures: $T_{\rm S}=310$ K (blue squares), $T_{\rm S}=479$ K 
(green circles), and  $T_{\rm S}=605$ K (red triangles). The
environment temperature is maintained at $T_{\rm E}=310$ K. The error
bars represent the total uncertainty (statistical and systematic)
of the measurement. (b) Average values $\gamma$ over the trap
center-surface separations of 7.0, 7.5, and 8.0 $\mu{\rm m}$,  plotted versus
substrate temperature. It is evident a clear increase in strength
of the atom-surface interaction for elevated temperatures. The solid theory curve
represents the non-equilibrium effect, while the dash-dotted theory curve 
represents the
case of equal temperatures.
}
\label{fig:Cornell4}
\end{SCfigure}

The experiment of Ref.\cite{Harber05} did not reach the accuracy to 
discriminate between the theoretical predictions
at $T=0\,{\rm K}$ and the $T=300\,{\rm K}$,
and a clear evidence of 
thermal effects was still missing. In this experiment there was no room to 
increase the temperature of the surface: at high temperatures 
atoms thermally desorb from the walls of the cell,
the vacuum in the cell degrades, resulting in the impossibility to produce a BEC. To 
overcome this experimental limitation a new configuration was studied, where only 
the surface temperature was increased: the quality of the vacuum was not affected
because of the relatively small size of the substrate. 
The non-equilibrium theory of atom-surface 
interactions in this system was developed in 
Refs.~\cite{Antezza05,Antezza06,Pitaevskii06,AntezzaPhD06, Antezza08}, 
as outlined in Sec.\ref{s:non-equilibrium}.
It predicts new qualitative and quantitative effects 
with respect to global equilibrium that are illustrated in
Fig.~\ref{fig:Force_NEQ}(right).
The experimental measurement has been achieved in 2007 \cite{Obrecht07}
and remains up to
now the only one that has detected thermal effects of the electromagnetic
dispersion interactions in this range of distances.
A sketch of the experimental apparatus is given in 
Fig.~\ref{fig:Cornell3}, the experimental results in Fig.~\ref{fig:Cornell4}.

%\begin{figure}[htb]
%\begin{center}
%%\includegraphics[width=8cm,clip=]{Figura_cm_Block_neq.eps}
%\caption{Theoretical prediction of the effect of the atom-surface force on trapped 
%atomic gases. (a) relative shift of the center-of-mass oscillation of a trapped 
%atomic BEC (gas radius of $2.5\mu{\rm m}$) of $^{87}$Rb atoms close to a sapphire 
%substrate. (b) relative shift of the Block oscillations of a Fermi gas trapped of $^
%{40}$K atoms in an optical lattice close to a sapphire substrate. This figure taken 
%from fig.2 of \cite{Antezza06}. \label{fig:CM_BLOCK_NEQ}}
%\end{center}
%\end{figure}

\subsubsection{Outlook}

Precision measurements of the atom-surface interaction may shed light
on the ongoing discussion about the temperature dependence of 
dispersion interactions with media that show absorption, like any conducting
medium. It has been pointed out 
by Klimchitskaya and Mostepanenko
\cite{Klimchitskaya08c}
that if the small, but nonzero conductivity
of the glass surface in the Cornell experiment \cite{Harber05,Obrecht07} had
been taken into account, the Lifshitz-Keesom tail would involve an infinite
static dielectric function, and hence deviate from a dielectric medium where
$1 < \varepsilon ( 0 ) < \infty$. This theoretical prediction would also be inconsistent
with the data. This issue is related to similar problems that arise in the 
macroscopic Casimir interaction, see 
Refs.\cite{Brown-Hayes06,Klimchitskaya09,Milton09a} for reviews. In the atom-surface
case, Pitaevskii has pointed out that a smooth crossover from a metal to a dielectric
is obtained within a non-local description that takes into account electric screening
in the surface (wave-vector-dependent dielectric function $\varepsilon( {\bf k},
\omega )$) \cite{Pitaevskii08} which Ref.\cite{Klimchitskaya08c} did not include. 

New interesting experimental proposals have been presented in order to measure 
the atom-surface force with higher accuracy, essentially based on interferometric 
techniques. All of them deal with atoms trapped in a periodic lattice 
made by laser beams (``optical lattice'' \cite{Grynberg01, Bloch08}),
and placed close to a substrate.
Gradients in the potential across the lattice can be detected
with coherent superposition states of atoms over adjacent lattice sites 
\cite{Wolf07}. These gradients also induce Bloch oscillations through
the reciprocal space of the lattice: if $\hbar q$ is the width of the Brillouin
zone, the period $\tau_B$ of
the Bloch oscillations is \cite{Carusotto05}
\begin{equation}
	\frac{ 1 }{ \tau_B } = \frac{ \overline{ - \partial_L U }} { \hbar q }
	\label{eq:Bloch-period}~.
\end{equation}
%(see fig.\ref{fig:} for the theoretical calculation of 
%the effect, taken from \cite{Antezza06}),  
where the average (overbar) is over the cloud size in the lattice. The
atom-surface interaction would, in fact, only shift the Bloch period
by a relative amount of $10^{-4} \ldots 10^{-3}$ if the main force is 
gravity and the atoms are at a distance $L \approx 5\,\mu{\rm m}$
\cite{Antezza06}.
Distance-dependent shifts in atomic clock frequencies have also been proposed
\cite{Derevianko09}. They arise from the differential energy shift of the two
atomic states which are related to the difference in polarizabilities. 
Finally, a corrugated surface produces a periodic Casimir-Polder potential that
manifests itself by a band gap in the dispersion relation of the elementary
excitations of the BEC (Bogoliubov modes). The spectrum of these modes 
is characterized by a dynamic structure factor that can be detected by the
Bragg scattering of a pair of laser beams \cite{Moreno10}.

\section*{Aknowledgments}

The authors would like to thank H. Haakh for fruitful discussions and comments, and R. Behunin for a critical reading. 

%% \clearpage
%
%
%%\bibliographystyle{atchip}
%\bibliography{BookBibliography}
%\bibliographystyle{/Users/nabu/Documents/Mydocs/Lavoro/bibliography/bibstyle/prstytitle}
%%\newcommand{\mybibpath}{/Users/carstenh/Work/Dropbox/Biblio/}
%%%% 
%%\bibliography{\mybibpath journals,%
%%bookCasimirComplete,%,
%%\mybibpath bib-ac,\mybibpath bib-dh,%
%%\mybibpath bib-io,\mybibpath bib-pz,%
%%\mybibpath bib-2004,\mybibpath bib-2005,%
%%\mybibpath bib-2006,\mybibpath bib-2007,%
%%\mybibpath bib-2008,\mybibpath bib-2009,%
%%\mybibpath bib-2010}%

\end{document}